\documentclass[twocolumn,showkeys,showpacs,nofootinbib]{revtex4}

\begin{document}


\title{Translations and dynamics}


\author{Romualdo Tresguerres}
\email[]{romualdotresguerres@yahoo.es}
\affiliation{Instituto de Matem\'aticas y F\'isica Fundamental\\
Consejo Superior de Investigaciones Cient\'ificas\\ Serrano 113
bis, 28006 Madrid, SPAIN}

\date{\today}

\begin{abstract}
We analyze the role played by local translational symmetry in the context of gauge theories of fundamental interactions. Translational connections and fields are introduced, with special attention being paid to their universal coupling to other variables, as well as to their contributions to field equations and to conserved quantities.
\end{abstract}

\maketitle

\section{Introduction}
Translational invariance is the main symmetry underlying Classical Mechanics, being responsible for linear momentum conservation, and thus for the law of action and reaction and for inertial motion. Therefore, it is amazing to realize the nearly irrelevant role, if any, assigned to such symmetry in other dynamical contexts, in particular in those concerned with basic interactions, such as General Relativity and gauge theories, where generalized forms of momentum occur.

Global spacetime translations, as a constitutive part of the Poincar\'e group, are certainly recognized as essential for the spacetime conception of Special Relativity. But as far as General Relativity makes appearance, general covariance disguises the meaning of local translations. Furthermore, local translational symmetry is usually ignored in the context of gauge theories, with few exceptions provided by a particular approach to gravity based on local spacetime groups such as the Poincar\'e or the affine one \cite{Kibble:1961ba} \cite{Trautman:1970cy} \cite{Cho:1975dh} \cite{Hehl:1974cn} \cite{Hehl:1976kj} \cite{Hehl:1979xk} \cite{Hehl:1995ue} \cite{Lord:1986xi} \cite{Lord:1987uq} \cite{Julve:1994bh} \cite{Tresguerres:2000qn} \cite{Tiemblo:2005js}. We conclude that the central role of translational invariance as a foundational principle remains far from being universally recognized, when it is not even explicitly refused by claiming it to be necessarily violated \cite{Petti:2006ue}.

The aim of the present paper is to uncover the hidden presence of local translational symmetry in the context of gauge theories. This will be achieved by considering the gauging of a spacetime group together with an internal group, exploiting the virtualities of certain suitable translational variables introduced in previous papers \cite{Julve:1994bh} \cite{Tresguerres:2000qn} \cite{Tiemblo:2005js} \cite{Tresguerres:2002uh}. For the sake of simplicity we choose Poincar\'e $\otimes$ $U(1)\,$ as the gauge group, with electrodynamics taken as a characteristic representative of general Yang-Mills theories. However, the interplay we are going to show, concerning the universal coupling of the translational variables to gauge potentials and fields of the remaining symmetries, is easily generalizable to any internal group, so that all our results are applicable to the whole Standard Model by considering Poincar\'e $\otimes$ $SU(3)\otimes SU(2)\otimes U(1)\,$; a simple task which is left to the reader. With special care in explicitly displaying the role played by translations, we will begin closely following the steps of Hehl et al. \cite{Hehl:1995ue} to develop a Lagrangian formalism giving rise to the field equations and to the Noether identities connected to the gauge symmetry. Then, an apparent digression on the rudiments of a Hamiltonian approach leads us to the identification of a well behaved --automatically conserved-- energy current 3-form related to the translational variables.

The paper is organized as follows. In Section II we recall the significance of translations for Newtonian Mechanics, showing the main lines of the way to go on. In Section III we discuss an exterior calculus reformulation of the standard variational principles. In IV we derive the field equations, and in V the Noether identities. In VI a Hamiltonian-like 3-form is introduced, and a definition of a conserved energy current--different from the (vanishing) Hamiltonian one-- is suggested. In order to illustrate the previous results with more familiar formulas, in VII we derive spacetime relations \cite{Hehl-and-Obukhov} between excitations and field strengths generalizing the electromagnetic case (\ref{emmom}), using several common Lagrangian pieces for matter and for fundamental interactions. In VIII we outline a Hamiltonian formalism containing a generalized translational Gauss law as the constraint acting as generator of translations. The paper ends with several final remarks in IX and with the Conclusions. However, we still leave for the appendices some related comments on the geometrical and kinematical interpretation of the formalism.
\newpage

\section{Global translations in Newtonian dynamics}

\subsection{Laws of motion}

In Classical Mechanics, linear momentum conservation, as derived from global space translations with the help of Noether's theorem, constitutes the ground where Newton's motion equations rest on. Actually, the law of inertia expresses conservation of the momentum of an isolated particle, while the law of action and reaction is the necessary and sufficient condition for momentum conservation of a system consisting of two particles. As for the forces introduced by the second law, they are suitably defined as quantities measuring the mutually compensating change induced on the momenta of the individual bodies, in such a way that conservation of the total linear momentum is guaranteed.

The fundamental role played by translations in Newtonian dynamics is explicitly shown by considering a system constituted by two particles, characterized by a Lagrangian depending on their positions and velocities, that is ${\cal L}={\cal L}\,( x_{_1}^a\,,x_{_2}^a\,;\dot{x}_{_1}^a\,,\dot{x}_{_2}^a\,)\,$ (where the dot denotes as usual derivation with respect to the time parameter $t\,$), being the linear momenta of the particles respectively defined as $p^{_{(1)}}_a :={\partial {\cal L}}/{\partial{\dot{x}_{_1}^a}}\,$,
$p^{_{(2)}}_a :={\partial {\cal L}}/{\partial{\dot{x}_{_2}^a}}\,$. A generic variation of such Lagrangian yields
\begin{eqnarray}
\delta {\cal L}&=&\delta x_{_1}^a\,\Bigl(\,{{\partial{\cal L}}\over{\partial{x_{_1}^a}}}-{{dp^{_{(1)}}_a}\over{dt}}\,\Bigr)
+\delta x_{_2}^a\,\Bigl(\,{{\partial{\cal L}}\over{\partial{x_{_2}^a}}}-{{dp^{_{(2)}}_a}\over{dt}}\,\Bigr)\nonumber\\
&&+{d\over{dt}}\Bigl(\delta x_{_1}^a\,p^{_{(1)}}_a +\delta
x_{_2}^a\,p^{_{(2)}}_a\Bigr)\,.\label{CM1}
\end{eqnarray}
Assuming the derived term in (\ref{CM1}) to vanish at the integration limits, the principle of least action requiring the action $S=\int{\cal L}\,dt\,$ to be extremal gives rise to the motion equations
\begin{equation}
{{\partial{\cal L}}\over{\partial{x_{_1}^a}}}
-{{dp^{_{(1)}}_a}\over{dt}}=0 \,,\qquad {{\partial{\cal L}}\over{\partial{x_{_2}^a}}}-{{dp^{_{(2)}}_a}\over{dt}}=0
\,,\label{CM2}
\end{equation}
where the gradients in (\ref{CM2}) are identifiable as forces, illustrating Newton's second motion equation for conservative forces deriving from a potential.

Now we return back to (\ref{CM1}) presupposing the motion equations (\ref{CM2}) to hold, and instead of a general variation, we perform a rigid displacement of the whole system. That is, we consider a translational group variation characterized by the constant parameters $\epsilon ^a$. Since we are dealing with global transformations, we assume the variation to be simultaneously well defined at distant places, being the same for both separated position variables $x_{_1}^a$ and $x_{_2}^a\,$, moved simultaneously as $\delta  x_{_1}^a =\delta  x_{_2}^a =-\epsilon ^a\,$. So one gets
\begin{equation}
\delta {\cal L}= {d\over{dt}}\Bigl(\delta  x_{_1}^a\,p^{_{(1)}}_a +\delta
x_{_2}^a\,p^{_{(2)}}_a\Bigr) =-\epsilon
^a\,{d\over{dt}}\Bigl(\,p^{_{(1)}}_a +p^{_{(2)}}_a\Bigr)\,.\label{CM3}
\end{equation}
From (\ref{CM3}) we read out that invariance under translations requires the conservation of linear momentum
\begin{equation}
{d\over{dt}}\Bigl(\,p^{_{(1)}}_a
+p^{_{(2)}}_a\Bigr)=0\,,\label{CM4}
\end{equation}
which is a condition not contained in (\ref{CM2}). Actually, by replacing (\ref{CM2}) in (\ref{CM4}), we get the law of action and reaction
\begin{equation}
{{\partial{\cal L}}\over{\partial{x_{_1}^a}}}+ {{\partial{\cal L}}\over{\partial{x_{_2}^a}}}=0\,,\label{CM5}
\end{equation}
affecting the forces appearing in (\ref{CM2}). Eq.(\ref{CM5}) is a direct consequence of translational invariance, implying the Lagrangian dependence on the individual positions $x_{_1}^a\,$, $x_{_2}^a$ to appear as dependence on the relative position $x_{_1}^a-x_{_2}^a$ of both particles, that is ${\cal L}(x_{_1}^a\,,x_{_2}^a\,;...)={\cal L}(x_{_1}^a-x_{_2}^a\,;...)\,$. Besides Newton's second law (\ref{CM2}) and third law (\ref{CM5}), one also obtains the first one by considering a system consisting of a single particle. Being the latter isolated in the universe, no forces are present and (\ref{CM4}) reduces to ${dp^{_{(1)}}_a}/{dt} =0\,$, expressing the principle of inertia concerning a single particle.

Historically, Descartes was pioneer in postulating a rough (scalar) version of momentum conservation (\,say $p^{_{(1)}}+p^{_{(2)}}=const.$\,) in the context of contact interactions as occurring in collisions. The improved continuous (vector) formulation (\ref{CM4}) of this principle --derivable, as already shown, from local translational invariance-- suggests the introduction of Newton's forces as quantitatively reflecting the soft changes of the momenta, see (\ref{CM2}), thus being interpretable as measures of non-contact interactions. Since, according to (\ref{CM4}), mutually compensating changes of momenta occur simultaneously at separated places as a result of global space-translational symmetry, instantaneous action at a distance as admitted in Newtonian mechanics manifests itself as a byproduct of such symmetry.

\subsection{The guiding principles}

The previous derivation of Newton's laws is based on a variational principle together with a symmetry principle, being both generalizable as powerful form-giving instruments underlying diverse dynamical formulations. However, they don't contain the complete physical information. Indeed, a third non actually existing principle would be necessary to entirely deduce empirically meaningful equations, both in the Newtonian as much as in the gauge-theoretical framework. When applied to the former classical example, the lacking principle should be responsible for justifying a Lagrangian
\begin{equation}
{\cal L}={1\over 2}\,m_{_1}\dot{x}_{_1}^2
+{1\over 2}\,m_{_2}\dot{x}_{_2}^2-V(x_{_1}^a -x_{_2}^a )\,,\label{Newton1}
\end{equation}
(where the potential $V$ could also be specified), allowing to go beyond the mere form of equations (\ref{CM2}) and (\ref{CM4}) by yielding the empirically relevant quantities $p^a_{_{(1)}} =m_{_1}\dot{x}_{_1}^a\,$ and $p^a_{_{(2)}} =m_{_2}\dot{x}_{_2}^a\,$ characteristic for Classical Mechanics. Regarding gauge theories, the lacking principle would be expected to provide a criterium to establish the form of the Lagrangian giving rise for instance to suitable spacetime relations of the Maxwell-Lorentz type (\ref{emmom}), determining the generalized excitations studied in Section VII. Since we don't have such a third principle, we are limited to adopt several Lagrangian pieces as established by experience.

Of course, we had avoided effort by having directly taken (\ref{Newton1}) as the starting point to derive the classical dynamical equations, since this Lagrangian resumes all the information discussed previously. However, by doing so we had lost the possibility of studying separately the contributions to the conformation of physical laws coming from each of the different principles invoked. Actually, from our treatment of the example of Newtonian Mechanics, we read out a general scheme to be kept in mind for what follows, consisting of three steps.

1.-- First we consider the {\it least} action (in fact, the {\it extremal} action) variational principle giving rise to the field equations in terms of quantities to be determined. The application of the principle does not require to know the particular form of the Lagrangian. One merely has to choose the dynamical variables, taking the Lagrangian to be a functional of them and of their (first) derivatives. If symmetry conditions are still not taken into account, the resulting field equations are trivially non covariant, see (\ref{bruttofieldeqs1})--(\ref{bruttofieldeqs6}) below.

2.-- Covariance is a consequence of the symmetry principle requiring the field equations to be compatible with invariance of the action under transformations of a particular symmetry group, see (\ref{covfieldeq1})--(\ref{covfieldeq3}) below. Depending on the group parameters being constant quantities or not, symmetries are global or local, both relating to conservation laws through Noether's theorem. (The symmetry principle in its local form is the gauge principle.)

3.-- Finally, from the lacking third principle we would expect a guide for establishing the fundamental spacetime relations analogous to the Maxwell-Lorentz electromagnetic one (\ref{emmom}). As a succedaneum of such principle, we take as guaranteed by a long experience the well established form of the Lagrangians of Dirac matter and electromagnetism, while for gravity we choose from the literature \cite{Obukhov:2006ge} a reasonable generalization (\ref{gravlagr}) of the Hilbert-Einstein Lagrangian for gravity, including quadratic terms in the irreducible pieces of torsion and curvature, constituting a tentative form to be adjusted by fixing certain parameters. See Section VII.

\section{Variational treatments of the action}

We use a formalism based on exterior calculus \cite{Hehl:1995ue}, with differential forms playing the role of dynamical variables. The fundamental kinds of objects involved in gauge theories consist of connections (\,1-forms\,) and fields (\,0-forms\,), both of them (denoted generically as $Q$ with all indices suppressed for simplicity) being fiber bundle constitutive elements. In terms of $Q$ and of their exact differentials, we build the Lagrangian density 4-form as a functional $L\,( Q\,,dQ\,)$, whose integral on a compact four-dimensional region  ${\cal D}$ of the bundle base space $M$ constitutes the action
\begin{equation}
S:=\int _{{\cal D}}L\,( Q\,,dQ\,)\,.\label{action}
\end{equation}
The bundle structure provides a geometrical background for different variational and symmetry considerations. In fact, in a bundle, two mutually orthogonal sectors exist, being the fibers regarded as vertical while the base space is conventionally taken as horizontal. Accordingly, two different kinds of variations are distinguished, depending on whether one moves vertically --by keeping fixed the integration domain--, or one alternatively considers horizontal displacements to neighboring integration regions of the base space \cite{Hecht:1993}. Each of these main categories of variations can be approached in different manners. So, besides generic vertical variations of the fields required to leave (\ref{action}) stationary in virtue of the principle of least action giving rise to the Euler-Lagrange equations, one has to consider the important particular case of vertical automorphisms along fibers, providing the bundle interpretation of gauge transformations. On the other hand, the action is not required to be left invariant under horizontal motions in order to derive dynamical laws. Nevertheless, such invariance can actually occur. For instance, displacements along base space paths generated by Killing vectors play the role of base space symmetry transformations.

\subsection{Vertical variations}

We first consider variations of (\ref{action}) affecting the
variables $Q$ (transforming $Q$ into $\hat{Q}\,$, say\,) while leaving the base space integration domain ${\cal D}$ untouched, so that
\begin{equation}
\delta\,S:=\int _{\cal D}\delta
L\,,\label{genericvar}
\end{equation}
where the integrated variation is to be understood as the infinitesimal limit of the difference $L\,(\hat{Q}\,,d\hat{Q}\,)-L\,( Q\,,dQ\,)$. In view of the functional dependence of (\ref{action}), the chain rule yields
\begin{equation}
\delta L =\delta Q\wedge {{\partial L}\over{\partial Q}} +\delta
dQ\wedge {{\partial L}\over{\partial dQ}}\,,\label{varlag1}
\end{equation}
which, being $[\,\delta\,,d\,]=0$, is trivially brought to the form
\begin{equation}
\delta L = \delta Q\wedge\left[\,\,{{\partial L}\over{\partial
Q}} -(-1)^p\,\,d\left({{\partial L}\over{\partial
dQ}}\right)\,\right] + d\,\left(\delta Q\wedge {{\partial L}\over{\partial
dQ}}\,\right)\,,\label{varlag4}
\end{equation}
analogous to (\ref{CM1}) with $p$ corresponding to the degree of the $p$-form $Q$. Variations of the action as given by (\ref{varlag4}) reveal to be useful to formalize both principles 1 and 2 of Section II B. These complementary impositions of vertical invariance of the action mainly differ from each other in the kind of field transformations considered in each case --namely generic variations {\it versus} group variations-- as much as in the dissimilar treatments applied to the exact term in (\ref{varlag4}).

On the one hand, the variational principle of extremal action demands the vertical invariance of the action (\ref{action}) by simultaneously imposing boundary conditions. According to Stokes' theorem\footnote{The Stokes theorem establishes
$$\int _{{\cal D}}d\,\omega = \int _{\partial {\cal D}}\omega\,,$$
being \(\omega\) a $p$-form on the $(p+1)$-dimensional compact integration domain ${\cal D}$ of the manifold $M$, with boundary $\partial {\cal D}$.}, (\ref{genericvar}) with (\ref{varlag4}) yields
\begin{equation}
\delta\,S =\int _{\cal D}\delta Q\wedge \left[\,\,{{\partial L}\over{\partial
Q}} -(-1)^p\,\,d\left({{\partial L}\over{\partial
dQ}}\right)\,\right] +\int _{\partial {\cal D}}\delta Q\wedge {{\partial
L}\over{\partial dQ}}\,.\label{explicitgenericvar}
\end{equation}
Stationarity of the action is imposed inside the integration domain ${\cal D}$ for generic variations $\delta Q$, arbitrary everywhere but at the integration boundary, where they are fixed (like the borders of a vibrating membrane, say) so as to cancel out the hypersurface term . In this way we derive the Euler-Lagrang equations
\begin{equation}
{{\partial L}\over{\partial Q}}
-(-1)^p\,\,d\left({{\partial L}\over{\partial dQ}}\right) =0
\,,\label{fieldeq}
\end{equation}
generalizing (\ref{CM2}). On the other hand, one can attend to the symmetry principle by considering gauge group transformations instead of arbitrary variations by taking $\delta Q$ as describing vertical automorphisms on the bundle \cite{Bleecker}. By requiring the field equations (\ref{fieldeq}) still to hold, the vanishing of (\ref{varlag4}) then reduces to that of the exact term, yielding the symmetry induced current conservation
\begin{equation}
d\,\left(\delta Q\wedge {{\partial L}\over{\partial dQ}}\,\right)
=0\,,\label{Noethercurrcons}
\end{equation}
according to Noether's theorem. (Compare with eq.(\ref{CM4}) of Newtonian Mechanics.) The new result (\ref{Noethercurrcons}) replaces the boundary condition by a symmetry requirement while keeping vertical invariance. We will show immediately how the consistence between (\ref{fieldeq}) and (\ref{Noethercurrcons}) causes the covariantization of the field equations by imposing suitable conditions on the partial derivatives ${{\partial L}\over{\partial Q}}$ occurring in (\ref{fieldeq}).

\subsection{Horizontal variations}

In addition, one can alternatively evaluate horizontal diffeomorphisms $f:M\rightarrow M$ acting on points $p\in M$ of the base space manifold \cite{Hecht:1993}. (A horizontal displacement on the base space of a bundle implies a displacement moving from fibers to fibers.) Using the notation $L\mid _p :=L\left[\,Q (p\,)\,,\,dQ(p\,)\,\right]\,$, and $L\mid _{f(p\,)}:=L\left[\,Q (f(p\,))\,,\,dQ(f(p\,))\,\right]\,$, we define the difference
\begin{equation}
\Delta _{\rm hor}\,S:=\int _{f({\cal D})} L\mid _{f(p\,)} -\int _{{\cal D}}
L\mid _p\label{hordiff}
\end{equation}
between the values of (\ref{action}) at domains displaced with respect to each other, where the notation $\Delta _{\rm hor}$ indicates that we are considering horizontal (base-space) diffeomorphisms. The pullback $f^*\,{}:\bigwedge T_{f(p)}^*\,M\rightarrow
\bigwedge T_p^*\,M$ induced by the diffeomorfism $f$ on differential forms $\omega$ satisfies $\int _{f({\cal D})}\omega = \int _{{\cal D}} f^*\omega\,$, thus allowing to rewrite the first term in the r.h.s. of (\ref{hordiff}) on the integration domain ${\cal D}\,$, so that it becomes comparable with the second one. By doing so while taking the diffeomorphism to depend on a parameter $s$ as $f_s$ and to be generated by a vector field ${\bf X}$, we find the horizontal variation (\ref{hordiff}) in the infinitesimal limit to reduce to
\begin{equation}
\delta _{\rm hor}\,S:=\int _{\cal D}\,\lim _{s\rightarrow 0}{1\over
s}\,\Bigl(\,f_s^*L\mid _{f_s(p\,)} - L\mid _p \,\Bigr)\,.\label{limhordiff}
\end{equation}
In view of the identity of the integrand with the standard definition of the Lie derivative \cite{Nakahara}, we finally get
\begin{equation}
\delta _{\rm hor}\,S = \int _{\cal D}{\it{l}}_{\bf x}
L\,.\label{L-Lie-der}
\end{equation}
The Lie derivative in (\ref{L-Lie-der}) measures the horizontal variation of the Lagrange density form along the vector field ${\bf X}$ on the base space. For arbitrary $p$-forms $\alpha$, the Lie derivative takes the explicit form
\begin{equation}
{\it{l}}_{\bf x}\alpha = X\rfloor d\,\alpha + d \left( X\rfloor\alpha\right)
\,.\label{Liederdef}
\end{equation}
A chain rule analogous to (\ref{varlag1}) holds for the Lagrangian Lie derivative in (\ref{L-Lie-der}) as
\begin{equation}
{\it{l}}_{\bf x} L ={\it{l}}_{\bf x} Q\wedge {{\partial L}\over{\partial Q}} +
{\it{l}}_{\bf x} dQ\wedge {{\partial L}\over{\partial
dQ}}\,.\label{Liechainrule}
\end{equation}
Only for certain vector fields ${\bf X}$ generating base space symmetries (Killing vectors), the Lie derivative (\ref{Liechainrule}) vanishes. In general, displacements on $M$ do not leave the Lagrangian form invariant, but they change it as ${\it{l}}_{\bf x} L\neq 0$. In view of (\ref{Liederdef}), we find the Lie derivative of the 4-form Lagrangian density to be ${\it{l}}_{\bf x} L :=d\,( X\rfloor L\,)\,$. Thus from (\ref{Liechainrule}), being $[\,{\it{l}}_{\bf x}\,,d\,]=0$, we find the identities
\begin{equation}
0=d\,\left[\,{\it{l}}_{\bf x} Q\wedge {{\partial L}\over{\partial dQ}}
-( X\rfloor L\,)\,\right] + {\it{l}}_{\bf x} Q\wedge {{\delta L}\over{\delta
Q}}\,,\label{idents}
\end{equation}
where we introduced the shorthand notation that we will use from now on for the variational derivative as appearing in (\ref{varlag4})--(\ref{fieldeq}), namely
\begin{equation}
{{\delta L}\over{\delta Q}}:= {{\partial L}\over{\partial Q}}
-(-1)^p\,\,d\left({{\partial L}\over{\partial dQ}}\right)
\,,\label{totalvar}
\end{equation}
whose vanishing means fulfillment of the field equations. Since we aren't going to consider base space symmetries, the non-vanishing r.h.s. of (\ref{L-Lie-der}) represents the effect of an admissible horizontal shift of the integration domain, while the {\it horizontal identities} (\ref{idents}) are merely a reformulation of the chain rule (\ref{Liechainrule}). However, provided the field equations hold --in view of vertical stationarity-- so that (\ref{totalvar}) vanishes, the {\it horizontal identities} (\ref{idents}) loose the last term, transforming into equations expressing the compatibility conditions between vertical invariance and horizontal displacements. The Noether type {\it identities} we are going to derive in Section V are of this kind .

\section{Gauging the Poincar\'e group times an internal symmetry}

The usually hidden role played by translations in gauge theories will be revealed by applying step by step the guiding principles presented in Section II B. We choose the Poincar\'e $\otimes$ $U(1)\,$ group, giving rise to a gauge theory of gravity and electromagnetism, because of its simplicity in considering together an internal and a spacetime symmetry including translations. But our results are applicable to other spacetime symmetry groups such as the affine group underlying metric-affine gravity \cite{Hehl:1995ue}, and to arbitrary internal groups yielding more general Yang-Mills theories such as the Standard Model or any other.

\subsection{The dynamical variables}

Regarding the particular treatment given in the present paper to translations, it may be clarifying to know that the author worked for a long time on nonlinear realizations of symmetries. It is in the context of nonlinear gauge approaches to several spacetime groups \cite{Julve:1994bh} \cite{Tresguerres:2000qn} \cite{Tiemblo:2005js} \cite{Tresguerres:2002uh} \cite{Lopez-Pinto:1997aw} that certain coordinate-like translational Goldstone fields $\xi ^\alpha$ occur, playing an important role in allowing the interpretation of tetrads as modified translative connections transforming as Lorentz covectors, thus making it possible to build Geometry entirely in gauge-theoretical (dynamical) terms.

In a previous paper \cite{Tresguerres:2002uh}, the author proposed a composite fiber bundle structure suitable to deal with nonlinear realizations of symmetries, and in particular with the gauge treatment of translations. The existence in such bundle of three mutually orthogonal sectors has as a consequence that translational fibers, although vertical when referred to the base space, may be regarded as defining an {\it intermediate base space} where other fibers are vertically attached to, as to a horizontal basis. Locality with respect to a given point $x$ of the genuine base space is compatible with displacements moving from a position $\xi ^\alpha (x)$ to a different one $\hat{\xi}^\alpha (x)\,$. So to say, the translational sector, characterized by the coordinate-like fields $\xi ^\alpha\,$, provides a dynamical spacetime background for the remaining bundle constituents.

Nevertheless, for what follows we don't need to support the coordinate-like fields theoretically on composite bundles. One can simply introduce such variables $\xi ^\alpha$, transforming as in (\ref{varcoordGoldstone}) below, regarding them as useful tools whose geometrical meaning as position vectors is discussed in Appendix B. In the following we will make an extensive use of these fields.

In order to deal with the Poincar\'e $\otimes$ $U(1)$ symmetry, we take as the fundamental dynamical variables $Q$ acting as arguments of (\ref{action}) the set
\begin{equation}
\{Q\}= \{\,\xi ^\alpha\,,\psi\,,\overline{\psi}\,, A\,,{\buildrel (T)\over{\Gamma
^\alpha}}\,,\Gamma ^{\alpha\beta}\}\,.\label{constfields}
\end{equation}
The quantities comprised in (\ref{constfields}) are either fields (0-forms) or connections (1-forms). Among them we recognize the previously discussed coordinate--like Goldstone fields $\xi ^\alpha$ and the matter fields chosen in particular to be Dirac spinors $\psi\,$ and $\overline{\psi}\,$ --all of them 0-forms-- and in addition we find the electromagnetic potential $A = dx^i A_i\,$, a translational connection ${\buildrel (T)\over{\Gamma ^\alpha}} =dx^i {\buildrel (T)\over{\Gamma _i^\alpha}}\,$, and the Lorentz connection $\Gamma ^{\alpha\beta} =dx^i \Gamma _i^{\alpha\beta}\,$, where the index $i$ refers to the underlying four-dimensional base space, while $\alpha = 0,1,2,3\,$ are anholonomic Lorentz indices, being the Lorentz connection antisymmetric in $\alpha\,,\beta\,$.

\subsection{Field equations and symmetry conditions}

The variation (\ref{varlag1}) of a Lagrangian density 4--form depending on variables (\ref{constfields}) and on their differentials reads
\begin{eqnarray}
\delta L =&&\delta\xi ^\alpha {{\partial L}\over{\partial\xi ^\alpha}}+\delta
d\xi ^\alpha\wedge {{\partial L}\over{\partial d\xi ^\alpha}} +\delta\overline{\psi}\,{{\partial
L}\over{\partial\overline{\psi}}}+\delta d\overline{\psi}\wedge
{{\partial L}\over{\partial d\overline{\psi}}}\nonumber\\
&&+ {{\partial
L}\over{\partial\psi}}\,\delta\psi +{{\partial L}\over{\partial
d\psi}}\wedge\delta d\psi +\delta A\wedge {{\partial
L}\over{\partial A}}+\delta d A\wedge {{\partial L}\over{\partial d A}}\nonumber\\
&&+\delta\Gamma _{^{(\,T)}}^\alpha\wedge {{\partial L}\over{\partial \Gamma
_{^{(\,T)}}^\alpha }}+\delta d\Gamma _{^{(\,T)}}^\alpha \wedge
{{\partial L}\over{\partial d\Gamma _{^{(\,T)}}^\alpha}}\nonumber\\
&&+\delta\Gamma ^{\alpha\beta}\wedge {{\partial L}\over{\partial
\Gamma ^{\alpha\beta}}}+\delta d\Gamma ^{\alpha\beta}\wedge
{{\partial L}\over{\partial d\Gamma ^{\alpha\beta}}}\,.\label{varphysLagr1}
\end{eqnarray}
According to the extremal action principle, the field equations (\ref{fieldeq}) are found to be
\begin{eqnarray}
{{\partial L}\over{\partial\xi ^\alpha}} -d\,{{\partial
L}\over{\partial d\xi ^\alpha}}&=&0\,,\label{bruttofieldeqs1}\\
{{\partial L}\over{\partial\overline{\psi}}} -d\,{{\partial L}\over{\partial
d\overline{\psi}}}&=&0\,,\label{bruttofieldeqs2}\\
{{\partial L}\over{\partial\psi}}+d\,{{\partial L}\over{\partial d\psi}}&=&0\,,\label{bruttofieldeqs3}\\
{{\partial L}\over{\partial A}} +d\,{{\partial L}\over{\partial
d A}}&=&0\,,\label{bruttofieldeqs4}\\
{{\partial L}\over{\partial \Gamma
_{^{(\,T)}}^\alpha }}+d\,{{\partial L}\over{\partial d\Gamma
_{^{(\,T)}}^\alpha}}&=&0\,,\label{bruttofieldeqs5}\\
{{\partial L}\over{\partial \Gamma
^{\alpha\beta}}}+d\,{{\partial L}\over{\partial d\Gamma
^{\alpha\beta}}}&=&0\,.\label{bruttofieldeqs6}
\end{eqnarray}
(Notice in particular the similitude between (\ref{bruttofieldeqs1}) and (\ref{CM2}).) On the other hand, according to the symmetry principle, the Noether conservation equation (\ref{Noethercurrcons}) takes the explicit form
\begin{eqnarray}
0=d &\Bigl[&\delta \xi ^\alpha\,{{\partial
L}\over{\partial d\xi ^\alpha}}+\delta\overline{\psi}\,{{\partial
L}\over{\partial d\overline{\psi}}} -{{\partial L}\over{\partial
d\psi}}\,\delta\psi +\delta A\wedge {{\partial L}\over{\partial d A}}\nonumber\\
&&+\delta {\buildrel (T)\over{\Gamma _\alpha}} \wedge {{\partial
L}\over{\partial d\Gamma _{^{(\,T)}}^\alpha}} +\delta\Gamma
^{\alpha\beta}\wedge {{\partial L}\over{\partial d\Gamma
^{\alpha\beta}}}\,\Bigr]\,.\label{conservedcurrs1}
\end{eqnarray}
For the Poincar\'e $\otimes$ $U(1)$ symmetry we are considering, the local group variations of the quantities (\ref{constfields}) are those of $U(1)$ together with the Poincar\'e ones as derived for instance in \cite{Tresguerres:2002uh}, that is
\begin{eqnarray}
\delta \xi ^\alpha &=&-\,\xi ^{\,\beta}\beta _\beta {}^\alpha
-\epsilon ^\alpha\,,\label{varcoordGoldstone}\\
\delta\psi &=&\left(\,i\lambda +i\beta ^{\alpha\beta}\sigma
_{\alpha\beta}\,\right)\,\psi\,,\label{varpsi}\\
\delta\overline{\psi}&=&-\,\overline{\psi}\,\left(\,i\lambda
+i\beta ^{\alpha\beta}\sigma _{\alpha\beta}\,
\right)\,,\label{varpsibar}\\
\delta A &=&-\,{1\over e}\,d\lambda\,,\label{varelectpot}\\
\delta {\buildrel (T)\over{\Gamma ^\alpha}} &=&-{\buildrel
(T)\over{\Gamma ^\beta}}\beta _\beta{}^\alpha +D\epsilon
^\alpha\,,\label{vartransconn}\\
\delta\Gamma _\alpha{}^\beta &=&D\beta _\alpha{}^\beta\,,\label{varlorconn}
\end{eqnarray}
with group parameters $\lambda (x)\,$, $\epsilon ^\alpha (x)\,$, $\beta ^{\alpha\beta}(x)\,$ (the latter being antisymmetric in $\alpha\,,\beta$) depending on the base space coordinates although not explicitly displayed\footnote{The covariant differentials in (\ref{vartransconn}) and (\ref{varlorconn}) are defined respectively as $$D\epsilon ^\alpha :=d\,\epsilon ^\alpha +\Gamma _\beta{}^\alpha\,\epsilon ^\beta \,,$$
$$D\beta ^{\alpha\beta} :=d\,\beta ^{\alpha\beta}
+\Gamma _\gamma{}^\alpha \beta ^{\gamma\beta}
+\Gamma _\gamma{}^\beta \beta ^{\alpha\gamma}\,.$$},
and being $\sigma _{\alpha\beta}$ the Lorentz generators in terms of Dirac gamma matrices. Intrinsic translations are not considered here, but the interested reader is referred to \cite{Tiemblo:2005sx} for a discussion on them. Rising and lowering of indices is performed by means of the constant Minkowski metric $o_{\alpha\beta}= diag(-+++)$ constituting the natural invariant of the Poincar\'e group. We remark the coordinate-like behavior of $\xi ^\alpha$ under transformations (\ref{varcoordGoldstone}), and we point out the transformation (\ref{vartransconn}) of ${\buildrel (T)\over{\Gamma ^\alpha}}$ as a connection, disqualifying it as a candidate to be identified as a tetrad.

Replacing in (\ref{conservedcurrs1}) the group variations (\ref{varcoordGoldstone})--(\ref{varlorconn}) we get
\begin{eqnarray}
&&0=d\,\Bigl\{\,{\lambda\over{e}}\,\left(\,J +d\,{{\partial
L}\over{\partial d A}}\,\right) -\epsilon
^\alpha\,\left(\,{{\partial L}\over{\partial d\xi ^\alpha}} +D
{{\partial L}\over{\partial d\Gamma
_{^{(\,T)}}^\alpha}}\,\right)\nonumber\\
&&-\beta ^{\alpha\beta}\,\left(\,\tau
_{\alpha\beta}+\xi _\alpha\,{{\partial L}\over{\partial d\xi
^\beta}}+{\buildrel (T)\over{\Gamma _\alpha}}\wedge {{\partial
L}\over{\partial d\Gamma _{^{(\,T)}}^\beta}}+ D {{\partial
L}\over{\partial d\Gamma ^{\alpha\beta}}}\,\right)
\,\Bigr\}\,,\nonumber\\
\label{conservedcurrs2}
\end{eqnarray}
where we introduced the definitions of electric current
\begin{equation}
J:=-ie\Bigl(\,\overline{\psi}\,{{\partial L}\over{\partial
d\overline{\psi}}} +{{\partial L}\over{\partial
d\psi}}\,\psi\,\Bigr)\,,\label{electcurr}
\end{equation}
and of spin current
\begin{equation}
\tau _{\alpha\beta}:=i\,\Bigl(\,\overline{\psi}\,\sigma
_{\alpha\beta}\,{{\partial L}\over{\partial d\overline{\psi}}}
+{{\partial L}\over{\partial d\psi}}\,\sigma
_{\alpha\beta}\,\psi\,\Bigr)\,.\label{spincurr}
\end{equation}
In order to deal with (\ref{conservedcurrs2}), we take from \cite{Hehl:1995ue} the property that a zero exact differential $d\,(\mu ^\alpha A_\alpha)=d\mu ^\alpha\wedge A_\alpha +\mu ^\alpha dA_\alpha =0$, with $\mu ^\alpha$ as much as $d\mu ^\alpha$ being pointwise arbitrary, implies the vanishing of both $A_\alpha$ and its differential. So from (\ref{conservedcurrs2}) we can derive the equations
\begin{eqnarray}
J +d\,{{\partial L}\over{\partial d A}} &=& 0\,,\label{coveq1}\\
{{\partial L}\over{\partial d\xi ^\alpha}} +D {{\partial
L}\over{\partial d\Gamma _{^{(\,T)}}^\alpha}} &=& 0\,,\label{coveq2}\\
\tau _{\alpha\beta} +\xi _{[\alpha }\,{{\partial
L}\over{\partial d\xi ^{\beta ]}}}+{\buildrel (T)\over{\Gamma
_{[\alpha}}}\wedge {{\partial L}\over{\partial d\Gamma
_{^{(\,T)}}^{\beta ]}}}+ D {{\partial L}\over{\partial d\Gamma
^{\alpha\beta}}} &=& 0\,,\hskip0.8cm\label{coveq3}
\end{eqnarray}
where the capital $D$ stands for the covariant differentials; see footnote 3. The compatibility between (\ref{coveq1})--(\ref{coveq3}) and the field equations (\ref{bruttofieldeqs4})--(\ref{bruttofieldeqs6}) requires the following consistence conditions to hold
\begin{eqnarray}
{{\partial L}\over{\partial A}}&=&J\,,\label{consistencecond1}\\
{{\partial L}\over{\partial \Gamma _{^{(\,T)}}^\alpha }} &=&
{{\partial L}\over{\partial d\xi ^\alpha}} -\Gamma
_\alpha{}^\beta\wedge {{\partial L}\over{\partial d\Gamma
_{^{(\,T)}}^\beta}}\,,\label{consistencecond2}\\
{{\partial L}\over{\partial \Gamma ^{\alpha\beta}}} &=& \tau
_{\alpha\beta} +\xi _{[\alpha }\,{{\partial L}\over{\partial d\xi
^{\beta ]}}}+{\buildrel (T)\over{\Gamma _{[\alpha}}}\wedge
{{\partial L}\over{\partial d\Gamma _{^{(\,T)}}^{\beta ]}}}\nonumber\\
&&\hskip0.6cm +2\,\Gamma _{[\alpha}{}^\gamma\wedge{{\partial L}\over{\partial
d\Gamma ^{\beta ]\gamma}}}\,.\label{consistencecond3}
\end{eqnarray}
Eq. (\ref{coveq3}) is not explicitly covariant, so that for the moment it is not evident that (\ref{consistencecond1})--(\ref{consistencecond3}), as derived with the help of the symmetry principle, just imply the covariantization of the field equations (\ref{bruttofieldeqs4})--(\ref{bruttofieldeqs6}) obtained previously. However, we are going to show that precisely that is the case. A further consistence condition follows from covariantly differentiating the covariant equation (\ref{coveq2}) to get
\begin{equation}
D {{\partial L}\over{\partial d\xi ^\alpha}}
-R_\alpha{}^\beta\wedge {{\partial L}\over{\partial d\Gamma
_{^{(\,T)}}^\beta}} =0\,,\label{derivedcoveq2}
\end{equation}
where $R_\alpha{}^\beta$ is the Lorentz curvature 2-form defined in (\ref{curvdef}). By comparing (\ref{derivedcoveq2}) with (\ref{bruttofieldeqs1}), we find
\begin{equation}
{{\partial L}\over{\partial\xi ^\alpha}} =\Gamma
_\alpha{}^\beta\wedge {{\partial L}\over{\partial d\xi ^\beta}}
+R_\alpha{}^\beta\wedge {{\partial L}\over{\partial d\Gamma
_{^{(\,T)}}^\beta}}\,.\label{consistencecond4}
\end{equation}
Notice that in (\ref{consistencecond4}) as much as in (\ref{consistencecond1})--(\ref{consistencecond3}), and in (\ref{CM5}) as well, it is the value of ${\partial L}/{\partial Q}$ the relevant one to enable covariance under the postulated symmetry. The covariantized form of (\ref{bruttofieldeqs1}) obtained by replacing the condition (\ref{consistencecond4}) is identical with (\ref{derivedcoveq2}) derived from (\ref{coveq2}). Thus (\ref{bruttofieldeqs1}) --its covariant version in fact-- results to be redundant.

\subsection{Fixing the notation}

The variation (\ref{varphysLagr1}) of the Lagrangian, together with the symmetry conditions (\ref{consistencecond1})--(\ref{consistencecond3})
and (\ref{consistencecond4}), yields what one would obtain by varying a Lagrangian already depending on covariant quantities, that is
\begin{eqnarray}
\delta L &=&\delta\vartheta ^\alpha\wedge {{\partial
L}\over{\partial d\xi ^\alpha}} +\delta \overline{D\psi}\wedge{{\partial L}\over{\partial
d\overline{\psi}}} +{{\partial L}\over{\partial
d\psi}}\wedge\delta D\psi\nonumber\\
&&+\delta\overline{\psi}\,\Bigl(\,{{\partial
L}\over{\partial\overline{\psi}}}-
{{\partial\overline{D\psi}}\over{\partial\overline{\psi}}}\wedge {{\partial
L}\over{\partial d\overline{\psi}}}\,\Bigr)\nonumber\\
&&+\Bigl(\,{{\partial
L}\over{\partial\psi}}-{{\partial L}\over{\partial
d\psi}}\wedge{{\partial D\psi}\over{\partial\psi}}\,\Bigr)\,\delta\psi\nonumber\\
&&+\delta F\wedge {{\partial L}\over{\partial d A}} +\delta T^\alpha\wedge {{\partial
L}\over{\partial d\Gamma _{^{(\,T)}}^\alpha}}\nonumber \\
&&+\delta R^{\alpha\beta}\wedge\left(\, {{\partial L}\over{\partial d\Gamma
^{\alpha\beta}}} -\xi _{[\alpha}\,{{\partial L}\over{\partial
d\Gamma _{^{(\,T)}}^{\beta ]}}}\,\right)\,,\label{varinvartlag}
\end{eqnarray}
where the original variables (\ref{constfields}) appear automatically rearranged into a number of Lorentz covariant objects defined in Appendix A, namely the tetrads $\vartheta ^\alpha $, the Lorentz $\otimes$ $U(1)$ covariant derivatives $D\psi\,$ and $\overline{D\psi}\,$ of the matter fields, the electromagnetic field strength $F$, the torsion $T^\alpha\,$ and the Lorentz curvature $R^{\alpha\beta}$.

In all these quantities, any vestige of explicit translational symmetry is absent, see (\ref{var-1})--(\ref{var-6}), explaining why translations, although genuinely present in the theory, become hidden. The ultimate reason for it is that the only original fields affected by translations according to (\ref{varcoordGoldstone})--(\ref{varlorconn}), namely $\xi ^\alpha\,$ and ${\buildrel (T)\over{\Gamma ^\alpha}}\,$, appear always joined together into the translation-invariant combination
\begin{equation}
\vartheta ^\alpha := D \xi ^\alpha +{\buildrel (T)\over{\Gamma
^\alpha}}\,,\label{tetrad}
\end{equation}
shown with more detail in (\ref{tetraddef}). Contrary to the original translative connection ${\buildrel (T)\over{\Gamma ^\alpha}}$, the modified one (\ref{tetrad}) transforms as a Lorentz covector, see (\ref{var-1}), making it possible to identify (\ref{tetrad}) as a tetrad, with a geometrical meaning compatible with its gauge-theoretical origin.

We further simplify the notation of several quantities also involved in (\ref{varinvartlag}). Firstly we define the canonical energy-momentum 3-form
\begin{equation}
\Pi _\alpha :={{\partial L}\over{\partial d\xi ^\alpha}}
\,,\label{definitions1(a)}
\end{equation}
resembling the classical definition $p_a :={\partial {\cal L}}/{\partial{\dot{x}^a}}\,$ of ordinary linear momentum. The symmetry condition (\ref{consistencecond2}) reveals a double character of (\ref{definitions1(a)}) by showing its equality --up to terms having to do with covariance-- with a translational current. It is in this second interpretation as a current that $\Pi _\alpha\,$ will behave as a source for gravitational fields, see (\ref{covfieldeq2}) below.

The ambiguity concerning the meaning of $\Pi _\alpha\,$ becomes increased by realizing, as we will do in Section V, that all fields of the theory contribute to this quantity. Decomposition (\ref{momentdecomp}) shows in fact that it consists of material, radiative and gravitational contributions, the double meaning affecting each of them. Usually it is illuminating to separate these different pieces from each other, mainly because matter currents $\Sigma _\alpha$ are naturally regarded as sources, while pure gravitational contributions $E_\alpha $ are of a different nature. But for the moment let us keep $\Pi _\alpha\,$ unified as a whole. By doing so the notation becomes simplified; and on the other hand, it is the complete $\Pi _\alpha$ that will play a role in the definition of the conserved energy current 3-form (\ref{energycurr}) to be defined in Section VI.

Otherwise, we follow Hehl's standard notation \cite{Hehl:1995ue}. Taking as a model the electromagnetic excitation 2-form
\begin{equation}
H:=-{{\partial L}\over{\partial d A}}\,,\label{definitions1(b)}
\end{equation}
(to be determined by the Maxwell-Lorentz spacetime relation (\ref{emmom})), we introduce its translative and Lorentzian gauge analogs, defined respectively as the 2-forms
\begin{equation}
H_\alpha :=-{{\partial L}\over{\partial d\Gamma
_{^{(\,T)}}^\alpha}}\,,\label{definitions1(c)}
\end{equation}
and
\begin{equation}
H_{\alpha\beta}:=-\left(\,{{\partial L}\over{\partial d\Gamma ^{\alpha\beta}}}- \xi _{ [\alpha}\,{{\partial L}\over{\partial d\Gamma _{(T)}^{\beta ]}}}\,\right)\,.\label{definition2}
\end{equation}
The second term in the r.h.s. of (\ref{definition2}) is due to the fact that, in view of (\ref{torsiondef}) with (\ref{tetraddef}), the torsion reads $T^\alpha := D\vartheta ^\alpha =D\,( D\,\xi ^\alpha +{\buildrel (T)\over{\Gamma
^\alpha}}\,)\,$, so that $\delta\,T^\alpha =\delta\,( R_\beta{}^\alpha \xi ^\beta + D{\buildrel (T)\over{\Gamma ^\alpha}}\,)$, having as a consequence the occurrence of a contribution to (\ref{definition2}) through the implicit dependence of $T^\alpha$ on $R_\beta{}^\alpha$. Comparison of (\ref{definitions1(a)}),(\ref{definitions1(c)}), (\ref{definition2}) with (\ref{varinvartlag}) reveals that
\begin{equation}
\Pi _\alpha :={{\partial L}\over{\partial \vartheta
^\alpha}}\,,\quad H_\alpha :=-{{\partial L}\over{\partial
T^\alpha}}\,,\quad H_{\alpha\beta}:=-\,{{\partial
L}\over{\partial R^{\alpha\beta}}}\,.\label{modificdefinitions}
\end{equation}
In terms of these objects we are going to rewrite (\ref{coveq1})--(\ref{coveq3}). However, first we have to reformulate the non explicitly covariant equation (\ref{coveq3}), making use of (\ref{tetrad}) and definitions (\ref{definitions1(a)}), (\ref{definitions1(c)}), (\ref{definition2}), as
\begin{equation}
DH_{\alpha\beta} +\vartheta _{[\alpha}\wedge H_{\beta ]} -\tau _{\alpha\beta} +\xi
_{\,[\alpha}\left(\,DH_{\beta ]} -\Pi _{\beta ]}\,\right) =0\,,\label{reformul}
\end{equation}
where the term in parentheses is merely (\ref{coveq2}), thus vanishing independently. So, the field equations (\ref{coveq1})--(\ref{coveq3}) take the form
\begin{eqnarray}
dH &=&J\,,\label{covfieldeq1} \\
DH_\alpha &=&\Pi _\alpha\,,\label{covfieldeq2}\\
DH_{\alpha\beta} +\vartheta _{[\alpha }\wedge H_{\beta ]}&=&\tau
_{\alpha\beta}\,.\label{covfieldeq3}
\end{eqnarray}
All of them are explicitly Lorentz covariant\footnote{The covariant differentials in
(\ref{covfieldeq2}) and (\ref{covfieldeq3}) are respectively defined as
$$DH_\alpha := dH_\alpha -\Gamma _\alpha{}^\beta\wedge H_\beta\,,$$
and
$$DH_{\alpha\beta} := dH_{\alpha\beta} -\Gamma _\alpha{}^\gamma\wedge H_{\gamma\beta}
-\Gamma _\beta{}^\gamma\wedge H_{\alpha\gamma}\,.$$}, while with respect to translations as much as to $U(1)$, they are invariant. In (\ref{covfieldeq1}) we recognize the Maxwell equations up to the explicit form of $H$ to be established in (\ref{emmom}). The fact that (\ref{covfieldeq2}) generalizes the gravitational Einstein equations is less evident, but see Section VII C. Both (\ref{covfieldeq2}) and (\ref{covfieldeq3}) reproduce the standard form established by Hehl et al. \cite{Hehl:1995ue}, with the main difference that in (\ref{covfieldeq2}) we do not separate the different pieces of $\Pi _\alpha$, as discussed above. The ambiguity derived from considering such a {\it source}, which is not a pure matter current -- as the electric current $J$ as well as the spin current $\tau _{\alpha\beta}$ are--, is compensated by the higher formal simplicity.

The redundant equation (\ref{derivedcoveq2}) constituting the covariantized version of (\ref{bruttofieldeqs1}) is immediately deducible from (\ref{covfieldeq2}) as
\begin{equation}
D \left( \Pi _\alpha -DH_\alpha\,\right) = D\Pi _\alpha +R_\alpha{}^\beta\wedge H_\beta =0\,.\label{nonnew}
\end{equation}
Thus, the simultaneous application of the variational principle yielding (\ref{bruttofieldeqs1})--(\ref{bruttofieldeqs6}), and of the symmetry principle, is summarized by the matter equations (\ref{bruttofieldeqs2})--(\ref{bruttofieldeqs3}) together with the covariant field equations (\ref{covfieldeq1})--(\ref{covfieldeq3}), the latter ones being associated respectively to $U(1)$, translations, and Lorentz symmetry.

\section{Noether identities}

In the present section we consider separately the different pieces \cite{Hehl:1995ue} into which one can meaningfully decompose the Poincar\'e $\otimes$ $U(1)$--invariant Lagrangian, that is
\begin{equation}
L=L^{\rm matt}+L^{\rm em}+L^{\rm gr}\,,\label{Lagrangedecomp}
\end{equation}
comprising on the one hand the material contribution $L^{\rm matt}(\,\vartheta
^\alpha\,,\psi\,,\overline{\psi}\,,D\psi\,,\overline{D\psi}\,)$,
plus an electromagnetic part $L^{\rm em}(\,\vartheta ^\alpha\,,\,F\,)\,$ and a pure gravitational constituent $L^{\rm gr}(\,\vartheta ^\alpha\,,\,T^\alpha\,,\,R_\alpha{}^\beta\,)$. Notice that the matter part of the Lagrangian depends basically on matter fields and their covariant derivatives, and the electromagnetic and gravitational pieces on the field strengths of the $U(1)$ and the Poincar\'e symmetry respectively. But not only. The universal $\vartheta ^\alpha$--dependence is also displayed everywhere. Actually, in Lagrangian pieces where the Hodge star operator $^*$ occurs, as it is the case for the physically realistic examples (\ref{Diraclagrang1}), (\ref{emlagrang1}) and (\ref{gravlagr}) to be considered later, this dependence is explicitly brought to light by the variational formula (\ref{dualvar}). Here we realize for the first time the (nonminimal) universal coupling of the translational variables comprised in the tetrad (\ref{tetrad}) to the remaining quantities of the theory, having as a consequence that all pieces in (\ref{Lagrangedecomp}) contribute to the energy-momentum (\ref{definitions1(a)}).

We are going to study the conditions for the vertical invariance of every separate part of (\ref{Lagrangedecomp}) under Poincar\'e $\otimes$ $U(1)$ gauge transformations (\ref{varcoordGoldstone})--(\ref{varlorconn}) --and the derived ones (\ref{var-1})--(\ref{var-6})--, as well as the compatibility conditions with the field equations of the horizontal displacements (\ref{Liechainrule}) of each independent Lagrangian piece along a generic vector field ${\bf X}$. We follow Hehl et al. \cite{Hehl:1995ue} in deriving simultaneously the Noether type conservation equations for matter currents, as much as the form of the different pieces
\begin{equation}
\Pi _\alpha =\Sigma ^{\rm matt}_\alpha +\Sigma ^{\rm em}_\alpha +E_\alpha
\label{momentdecomp}
\end{equation}
into which (\ref{definitions1(a)}) becomes decomposed consistently with (\ref{Lagrangedecomp}), with the obvious notation $\Sigma ^{\rm matt}_\alpha :=\partial L^{\rm matt}/{\partial d\xi ^\alpha}$, $\Sigma ^{\rm em}_\alpha :=\partial L^{\rm em}/{\partial d\xi ^\alpha}$ and $E_\alpha :=\partial L^{\rm gr}/{\partial d\xi ^\alpha}$, as read out from (\ref{definitions1(a)}) and (\ref{Lagrangedecomp}).

Let us start with the matter Lagrangian part $L^{\rm matt}\,$. For what follows, with the help of (\ref{consistencecond1})--(\ref{consistencecond3}) we identify the matter currents associated to the different symmetries as the derivatives of the matter Lagrangian with respect to the corresponding connection, as usual in gauge theories, that is
\begin{eqnarray}
J={{\partial L^{\rm matt}}\over{\partial A}}\,,\quad
\Sigma ^{\rm matt}_\alpha &=& {{\partial L^{\rm matt}}\over{\partial \Gamma _{^{(\,T)}}^\alpha }}\,,\nonumber\\
\tau _{\alpha\beta} +\xi _{[\alpha}\wedge\Sigma ^{\rm matt}_{\beta ]} &=& {{\partial L^{\rm matt}}\over{\partial \Gamma ^{\alpha\beta}}}\,.\label{mattcurrdefs}
\end{eqnarray}
Provided the field equations are fulfilled, the gauge transformations (\ref{varcoordGoldstone})--(\ref{varlorconn}) of $L^{\rm matt}$ yield
\begin{equation}
\delta L^{\rm matt} = {\lambda\over e}\,dJ -\beta ^{\alpha\beta}\,\Bigl(\,D\,\tau
_{\alpha\beta} +\vartheta _{[\alpha}\wedge\Sigma ^{\rm matt}_{\beta
]}\,\Bigr)\,.\label{varmattLagr}
\end{equation}
From the vanishing of (\ref{varmattLagr}), as required by its postulated Poincar\'e $\otimes$ $U(1)$ invariance, we read out first the conservation of the electric current (\ref{electcurr}), namely
\begin{equation}
dJ =0\,,\label{elcurrcons}
\end{equation}
a result which looks trivial in view of being also obtainable by merely differentiating (\ref{covfieldeq1}). Furthermore we get also the less simple conservation equation for the spin current
\begin{equation}
D\,\tau _{\alpha\beta} +\vartheta _{[\alpha}\wedge\Sigma
^{\rm matt}_{\beta ]}=0\,,\label{spincurrconserv}
\end{equation}
a result which is not {\it a priori} expected.

On the other hand, we consider a horizontal displacement of the matter part of the action, assuming simultaneously its vertical invariance by supposing the symmetry conditions (\ref{elcurrcons}) and (\ref{spincurrconserv}) to hold. The requirement of vertical invariance of the total action is also kept in mind, reflecting itself in the field equations. In this way we get new identities of the Noether type. For convenience, in our deduction we use (\ref{Liechainrule}) rather than the equivalent equation (\ref{idents}) due to the fact that the latter presents no calculational advantage in the present case. Indeed, the variational derivative term in (\ref{idents}) doesn't vanish for each Lagrangian piece separately, since field equations derive from the whole Lagrangian. The Lie derivative (\ref{Liechainrule}) of the matter piece of the Lagrangian satisfying the mentioned conditions expands as
\begin{eqnarray}
{\it{l}}_{\bf x} L^{\rm matt} &=&-X^\alpha \Bigl[\,D\,\Sigma ^{\rm matt}_\alpha
-(\,e_\alpha\rfloor T^\beta )\wedge\Sigma ^{\rm matt}_\beta\nonumber\\
&&\hskip0.8cm -(\,e_\alpha\rfloor R^{\beta\gamma}\,)\wedge\tau _{\beta\gamma}
-(\,e_\alpha\rfloor F\,)\wedge J\,\Bigr]\nonumber \\
&+&d\,\Bigl[\,X^\alpha\Sigma ^{\rm matt}_\alpha
+(\,X\rfloor\overline{D\psi}\,)\,{{\partial
L^{\rm matt}}\over{\partial d\overline{\psi}}}\nonumber\\
&&\hskip2.2cm -{{\partial L^{\rm matt}}\over{\partial d\psi}}\,(\,X\rfloor
D\psi\,)\,\Bigr]\,.\label{Liederivmatt}
\end{eqnarray}
Due to the fact that ${\it{l}}_{\bf x} L^{\rm matt} =d\,(X\rfloor L^{\rm matt})$, as read out from (\ref{Liederdef}) being $L^{\rm matt}$ a 4--form, (\ref{Liederivmatt}) can be reduced to the form $0=X^\alpha A_\alpha + d\,(X^\alpha B_\alpha\,) =X^\alpha (\,A_\alpha + d B_\alpha\,) +d\,X^\alpha B_\alpha$, so that --as before-- for pointwise arbitrary $X^\alpha$ and $dX ^\alpha$, the vanishing of both $A_\alpha$ and $B_\alpha$ follows \cite{Hehl:1995ue}, implying
\begin{equation}
D\,\Sigma ^{\rm matt}_\alpha =(\,e_\alpha\rfloor T^\beta )\wedge\Sigma
^{\rm matt}_\beta +(\,e_\alpha\rfloor R^{\beta\gamma}\,)\wedge\tau
_{\beta\gamma} +(\,e_\alpha\rfloor F\,)\wedge J\,,\label{sigmamattconserv}
\end{equation}
and
\begin{equation}
\Sigma ^{\rm matt}_\alpha =-(\,e_\alpha\rfloor\overline{D\psi}\,)\,{{\partial
L^{\rm matt}}\over{\partial d\overline{\psi}}} +{{\partial
L^{\rm matt}}\over{\partial d\psi}}\,(\,e_\alpha\rfloor D\psi\,) +
e_\alpha\rfloor L^{\rm matt}\,.\label{sigmamatt}
\end{equation}
Eq. (\ref{sigmamattconserv}) is a sort of force equation; see (\ref{reformsigmamattconserv}). Indeed, in the last of the similar terms entering the r.h.s. we recognize the ordinary Lorentz force involving the electromagnetic field strength and the electric current. The remaining pieces have the same structure, being built from the field strength and the matter current associated to translational and Lorentz symmetry respectively. On the other hand, (\ref{sigmamatt}) outlines the form of the matter part of (\ref{momentdecomp}). (Recalling the previously discussed ambiguity of energy-momentum, notice that $\Sigma ^{\rm matt}_\alpha\,$ in the r.h.s. of (\ref{sigmamattconserv}) behaves as one of the three kinds of matter currents present in the theory, while in the l.h.s. the same quantity is more naturally understood as matter momentum.)

Having finished our detailed study of the matter part of the Lagrangian, let us now briefly summarize the results obtained by proceeding analogously with the two remaining pieces in (\ref{Lagrangedecomp}). Regarding the electromagnetic Lagrangian constituent, its gauge transformation yields $\delta L^{\rm em} =-\beta ^{\alpha\beta}\,\vartheta
_{[\alpha}\wedge\Sigma ^{\rm em}_{\beta ]}\,$, so that its invariance implies the symmetry condition
\begin{equation}
\vartheta _{[\alpha}\wedge\Sigma ^{\rm em}_{\beta ]} =0\,.\label{Symem-emt}
\end{equation}
The equation analogous to (\ref{Liederivmatt}) yields
\begin{equation}
D\,\Sigma ^{\rm em}_\alpha =(\,e_\alpha\rfloor T^\beta )\wedge\Sigma
^{\rm em}_\beta -(\,e_\alpha\rfloor F\,)\wedge dH\,,\label{sigmaemconserv}
\end{equation}
as much as the form of the electromagnetic energy-momentum
\begin{equation}
\Sigma ^{\rm em}_\alpha =(\,e_\alpha\rfloor F\,)\wedge H + e_\alpha\rfloor L^{\rm em}\,.\label{sigmaem}
\end{equation}
Finally we consider the gravitational Lagrangian part. Its invariance condition
\begin{equation}
D\,\Bigl( DH_{\alpha\beta}
+\vartheta _{[\alpha }\wedge H_{\beta ]}\,\Bigr) +\vartheta _{[
\alpha}\wedge\Bigl( DH_{\beta ]} -E_{\beta
]}\,\Bigr)=0\,,\label{redund}
\end{equation}
turns out to be redundant with previous results since it can be immediately derived from the field equations (\ref{covfieldeq2}), (\ref{covfieldeq3}), together with (\ref{momentdecomp}), (\ref{spincurrconserv}) and (\ref{Symem-emt}). The (\ref{Liederivmatt})-- analogous equation gives rise to two different results. On the one hand, it yields
\begin{eqnarray}
&&D\,\Bigl( DH_\alpha -E_\alpha\,\Bigr) -(\,e_\alpha\rfloor T^\beta
)\wedge\Bigl( DH_\beta -E_\beta\,\Bigr)\nonumber\\
&&\hskip0.2cm -(\,e_\alpha\rfloor R^{\beta\gamma}\,)\wedge\Bigl( DH_{\beta\gamma}+\vartheta _{[\beta
}\wedge H_{\gamma ]}\,\Bigr)=0\,,\label{ealphaconserv}
\end{eqnarray}
which is also redundant, derivable from the field equations (\ref{covfieldeq1})--(\ref{covfieldeq3}) with (\ref{momentdecomp}), (\ref{sigmamattconserv}) and (\ref{sigmaemconserv}). On the other hand, it provides the form of the pure gravitational contribution to energy-momentum, namely
\begin{eqnarray}
E_\alpha =(\,e_\alpha\rfloor T^\beta )\wedge H_\beta
+(\,e_\alpha\rfloor R^{\beta\gamma}\,)\wedge H_{\beta\gamma}
+e_\alpha\rfloor L^{\rm gr}\,.\label{ealpha}
\end{eqnarray}
The total momentum (\ref{momentdecomp}) entering the field equation (\ref{covfieldeq2}) is found by performing the sum of (\ref{sigmamatt}), (\ref{sigmaem}) and (\ref{ealpha}) as
\begin{eqnarray}
\Pi _\alpha &=& -(\,e_\alpha\rfloor\overline{D\psi}\,)\,{{\partial
L}\over{\partial d\overline{\psi}}} +{{\partial L}\over{\partial
d\psi}}\,(\,e_\alpha\rfloor D\psi\,) +(\,e_\alpha\rfloor F\,)\wedge H \nonumber\\
&&+(\,e_\alpha\rfloor T^\beta )\wedge H_\beta +(\,e_\alpha\rfloor R^{\beta\gamma}\,)\wedge H_{\beta\gamma} +e_\alpha\rfloor L\,.\label{sum}
\end{eqnarray}
Written in this form, it will play a relevant role in what follows.

\section{Energy conservation}

In Section III B we introduced equation (\ref{idents}) governing horizontal displacements along arbitrary vector fields ${\bf X}$ on the base space, and we discussed the compatibility of such displacements with vertical invariance of the action (that is, with fulfillment of the field equations). Now we are going to particularize to the case of the prominent vector field $n$ characterized as follows. On the base space we introduce a 1-form $\omega$ satisfying the Frobenius' foliation condition $\omega\wedge d\omega =0$, whose general solution reads $\omega = N d\tau$. With the help of $\tau$ obtained in this way, taken to be --at least locally-- a monotone increasing variable, it becomes possible to parametrize nonintersecting 3-dimensional base space hypersurfaces. This justifies to regard $\tau$ as {\it parametric time}, while $N$ is the so called lapse function fixing a {\it time scale}. The vector $n$ acquires its {\it temporal} meaning through the condition $n\rfloor (N d\tau )=1$ relating it to the {\it parametric time} variable. The concept of {\it temporality} thus emerges from the foliation of the base space. (The same holds for {\it spatiality}, the latter however as a secondary result.) Indeed, in principle no a {\it time coordinate} is identifiable as such in the base space. It is through the foliation that {\it parametric time} $\tau $ appears, conforming its associated {\it parametric time vector field} $n$.

Horizontal displacements along $n$ given by the Lie derivative of any variable are to be understood as {\it parametric time evolution}. Being normal to the {\it spatial} hypersurfaces, the vector field $n$ is tangential to a congruence of worldlines. The direction defined by the {\it parametric time vector} on the base space allows to perform a decomposition \cite{Hehl-and-Obukhov} of any $p$-form $\alpha$ into two constituents, respectively longitudinal and transversal to $n$ as
\begin{equation}
\alpha = N d\tau\wedge\alpha _{\bot} +\underline{\alpha}\,,\label{foliat1}
\end{equation}
being the longitudinal component
\begin{equation}
\alpha _{\bot} :=n\rfloor\alpha\label{long-part}
\end{equation}
the projection of $\alpha$ along $n$, and the transversal part
\begin{equation}
\underline{\alpha}:=
n\rfloor (N d\tau\wedge\alpha\,)\,,\label{trans-part}
\end{equation}
an orthogonal projection on the spatial sheets.

The longitudinal part of the tetrad (\ref{tetrad}) will play a singular role due to the following formal reason. As discussed in Appendix B, one can introduce a vector basis $e_\alpha$ dual to the coframes (\ref{tetrad}) in the sense that $e_\alpha\rfloor\vartheta ^\beta =\delta _\alpha ^\beta$. Thus, being $\vartheta ^\alpha _{\bot}:=n\rfloor\vartheta ^\alpha $ according to (\ref{long-part}), one can express the vector field $n=n^i\partial _i$ alternatively as $n =\vartheta ^\alpha _{\bot} e_\alpha$. The fact that $n$ itself must be trivially time-like has its formal plasmation in the property $o_{\alpha\beta}\,\vartheta _{\bot}^\alpha\otimes\vartheta _{\bot}^\beta =-1$ read out from (\ref{quadrats}).

\subsection{Vanishing Hamiltonian-like 3-form}

Starting with the identity (\ref{idents}) valid for arbitrary vector fields, we apply it in particular to the {\it time vector} $n$. We do not perform here a complete foliation of the equations as we will do in Section VIII, where we will totally separate longitudinal and transversal parts from each other; but we make use of the notation (\ref{long-part}) as a convenient shorthand for quantities such as $n\rfloor Q := Q_{\bot}$ or $n\rfloor L =: L_{\bot}$. Analogously, applying (\ref{Liederdef}) particularized for the {\it parametric time vector} $n\,$, we denote
\begin{equation}
{\it{l}}_n Q := (n\rfloor d\,Q\,)+ d\,(n\rfloor Q\,) =:(
d\,Q\,)_{\bot}+ d\,Q_{\bot}\,,\label{EC04}
\end{equation}
compare with (\ref{longitderiv}). Using (\ref{EC04}) we rewrite (\ref{idents}) as
\begin{eqnarray}
0&=&d\,\left[\,Q_{\bot}\wedge{{\partial L}\over{\partial Q}} +
\left( d\,Q\right)_{\bot} \wedge {{\partial L}\over{\partial dQ}}
-L_{\bot} -Q_{\bot}\wedge {{\delta L}\over{\delta Q}}\,\,\right]\nonumber\\
&&\hskip0.4cm +{\it{l}}_n Q\wedge {{\delta L}\over{\delta Q}}\,.\label{EC05}
\end{eqnarray}
By defining the Hamiltonian-like 3-form
\begin{equation}
{\cal H} := \,Q_{\bot}\wedge{{\partial L}\over{\partial Q}} +
\left( d\,Q\right)_{\bot} \wedge {{\partial L}\over{\partial
dQ}}-L_{\bot} -Q_{\bot}\wedge {{\delta L}\over{\delta Q}}
\,,\label{EC06}
\end{equation}
eq. (\ref{EC05}) becomes
\begin{equation}
d{\cal H}+{\it{l}}_n Q\wedge {{\delta L}\over{\delta
Q}}=0\,.\label{EC07}
\end{equation}
Thus, provided the field equations (\ref{fieldeq}) hold, (\ref{EC07}) seems to yield a continuity equation $d{\cal H}=0\,$ affecting the quantity ${\cal H}$, the latter being a sort of energy current 3-form. Unfortunately, we are going to prove that such equation trivializes since ${\cal H}$ itself vanishes. To arrive at such conclusion, we evaluate (\ref{EC06}) explicitly for the variables (\ref{constfields}). Although not immediately evident, the first terms in the r.h.s. of (\ref{EC06}) can be rearranged into covariant expressions by replacing the symmetry conditions (\ref{consistencecond1})--(\ref{consistencecond3}), so that for fulfilled field equations, (\ref{EC06}) takes the gauge invariant form
\begin{eqnarray}
{\cal H} =&&\,\vartheta _{\bot}^\alpha \,{{\partial L}\over{\partial d\xi
^\alpha}}+\overline{{\cal \L\/}_n\psi}\,\,{{\partial
L}\over{\partial d\overline{\psi}}} -{{\partial L}\over{\partial
d\psi}}\,\,{\cal
\L\/}_n\psi\nonumber\\
&&+ F_{\bot}\wedge {{\partial L}\over{\partial dA}} +T_{\bot}^\alpha \wedge {{\partial
L}\over{\partial d\Gamma _{(T)}^\alpha}}\nonumber\\
&&+ R_{\bot}^{\alpha\beta}\wedge\left(\,{{\partial L}\over{\partial
d\Gamma ^{\alpha\beta}}}- \xi _{ [\alpha}\,{{\partial
L}\over{\partial d\Gamma
_{(T)}^{\beta ]}}}\,\right) -L_{\bot}\,,\label{EC12}
\end{eqnarray}
where we used definitions ${\cal\L\/}_n\psi := n\rfloor D\psi = (D\psi )_{\bot}\,$, compare with (\ref{covLiederiv2}), and $F_{\bot} := n\rfloor F\,$, etc.; see (\ref{long-part}). By returning back now to the previous result (\ref{sum}), contracting it with $\vartheta _{\bot}^\alpha $ and recalling that $n =\vartheta _{\bot}^\alpha e_\alpha\,$, we find
\begin{eqnarray}
0&=&\vartheta _{\bot}^\alpha \,\Pi _\alpha
+\overline{{\cal \L\/}_n\psi}\,\,{{\partial
L}\over{\partial d\overline{\psi}}} -{{\partial L}\over{\partial
d\psi}}\,\,{\cal \L\/}_n\psi - F_{\bot}\wedge H\nonumber\\
&&-T_{\bot}^\alpha\wedge H_\alpha -R_{\bot}^{\alpha\beta}\wedge H_{\alpha\beta} -L_{\bot}\,,\label{vanish-H}
\end{eqnarray}
revealing that (\ref{EC12}) reduces to zero. So, instead of a continuity equation $d{\cal H}=0\,$, we merely have a relation between the different terms in (\ref{EC12}), namely ${\cal H}=0\,$ or (\ref{vanish-H}). This result holds independently of the particular form of the Lagrangian, and it is in close relationship with the well known vanishing of any possible Hamiltonian of General Relativity.

\subsection{A well behaved energy current}

Since $d{\cal H}=0$ cannot play the role of a law of conservation of energy because of its triviality, we look for an alternative formulation of such law, if possible. At this respect, let us recall the singular role played by translational variables as compared with the remaining constituents of the theory, in the sense that $\xi ^\alpha$ and $\Gamma _{(T)}^\alpha\,$, confined together in the translation-invariant combination constituting the tetrad (\ref{tetrad}), couple to any other physical quantity (usually through the $\vartheta ^\alpha\,$--terms in (\ref{dualvar}), provided the Hodge dual operator occurs, as already mentioned). The universal coupling of translations compels information relative to any other quantity to become stored in the (translational) energy-momentum (\ref{sum}). Accordingly, in (\ref{vanish-H}) each contribution appears twice, so to say: once explicitly and once through $\Pi _\alpha\,$, with the result that the total sum cancels out. Having this fact in mind, we propose to identify in (\ref{vanish-H}) a meaningful expression to be defined as (translational) energy, balancing the joint amount of the remaining energy contributions. The possible energy candidate is expected to be conserved.

We find such a quantity effectively to exist, consisting in the energy current 3-form
\begin{equation}
\epsilon := -\left(\,\vartheta _{\bot}^\alpha\,\Pi _\alpha + D\vartheta _{\bot}^\alpha\wedge H_\alpha\,\right)\,,\label{energycurr}
\end{equation}
which in view of (\ref{covfieldeq2}) satisfies the nontrivial continuity equation
\begin{equation}
d\,\epsilon =0\,,\label{energyconserv}
\end{equation}
with the meaning of local conservation of energy. By rewriting (\ref{vanish-H}) in terms of (\ref{energycurr}) while taking into account (\ref{thetaLiederiv}), we get
\begin{eqnarray}
\epsilon =&&\overline{{\cal \L\/}_n\psi}\,\,{{\partial
L}\over{\partial d\overline{\psi}}} -{{\partial L}\over{\partial
d\psi}}\,\,{\cal \L\/}_n\psi -F_{\bot}\wedge H\nonumber\\
&&-{\cal \L\/}_n\vartheta ^\alpha \wedge H_\alpha -R_{\bot}^{\alpha\beta}\wedge H_{\alpha\beta} -L_{\bot}\,,\label{conservedenergy}
\end{eqnarray}
where the total --nonvanishing-- energy $\epsilon$ in the l.h.s. of (\ref{conservedenergy}) resumes the whole information concerning the remaining fields displayed in the r.h.s., as already commented.

We conclude that the singularity of the Hamiltonian (\ref{EC12}) is a consequence of the presence of translations, even if hidden, in the scheme. This result is unavoidable as far as gravitation is taken into account, since modified translational connections (\ref{tetrad}) --that is tetrads, or the Riemannian metric built from them-- are to be treated as dynamical variables, thus giving rise to the occurrence of a contribution (\ref{energycurr}) leading to the vanishing of ${\cal H}\,$. Notice that a nonvanishing Hamiltonian-like 3-form ${\cal H}$ with the ordinarily expected meaning of a nonvanishing energy current only would make sense in contexts where gravitational contributions (and thus translations) were disregarded.

According to (\ref{energycurr}) and taking the decomposition (\ref{momentdecomp}) into account, we introduce three different contributions to energy, namely
\begin{equation}
\epsilon =\epsilon ^{\rm matt}+\epsilon ^{\rm em}+\epsilon ^{\rm gr}
\,,\label{energydec}
\end{equation}
respectively defined as
\begin{eqnarray}
\epsilon ^{\rm matt} &:=& -\vartheta _{\bot}^\alpha\,\Sigma ^{\rm matt}_\alpha
\,,\label{mattenergy}\\
\epsilon ^{\rm em} &:=& -\vartheta _{\bot}^\alpha\,\Sigma ^{\rm em}_\alpha
\,,\label{emenergy}\\
\epsilon ^{\rm gr} &:=& -\left(\,\vartheta _{\bot}^\alpha\,E_\alpha + D\vartheta _{\bot}^\alpha\wedge H_\alpha\,\right)\,.\label{grenergy}
\end{eqnarray}
None of them is a conserved quantity. Actually, from (\ref{mattenergy}) with (\ref{sigmamattconserv}) we get for instance
\begin{equation}
d\,\epsilon ^{\rm matt} := -{\cal \L\/}_n\,\vartheta ^\alpha\wedge\Sigma ^{\rm matt}_\alpha -R_{\bot}^{\alpha\beta}\wedge\tau _{\alpha\beta} -F_{\bot}\wedge J\,.\label{mattender}
\end{equation}
The non zero r.h.s. of (\ref{mattender}) may be partially illuminated with the help of the remaining contributions to the total energy conservation (\ref{energyconserv}). Indeed,  (\ref{emenergy}) with (\ref{sigmaemconserv}) yields
\begin{eqnarray}
d\,\epsilon ^{\rm em} &:=& -{\cal \L\/}_n\,\vartheta ^\alpha\wedge\Sigma ^{\rm em}_\alpha + F_{\bot}\wedge dH\nonumber\\
&=& -{\cal \L\/}_n\,\vartheta ^\alpha\wedge\Sigma ^{\rm em}_\alpha + F_{\bot}\wedge J \,,\label{emender}
\end{eqnarray}
which we are going to compare in Section VII B with the well known electromagnetic energy conservation equation involving the Poynting vector and Joule's heat. (In the standard electromagnetic formulation, the first term in the r.h.s. of (\ref{emender}) is absent.) If desired, one can consider (\ref{emender}) as a modified form of the first law of Thermodynamics, an idea which is generalizable to the previous and the next case. For the gravitational energy (\ref{grenergy}) with (\ref{ealphaconserv}) we finally find
\begin{eqnarray}
d\,\epsilon ^{\rm gr} &:=& -{\cal \L\/}_n\,\vartheta ^\alpha\wedge\left(\,E_\alpha -DH_\alpha\right)\nonumber\\
&&+R_{\bot}^{\alpha\beta}\wedge \left(\,DH_{\alpha\beta} +\vartheta _{[\alpha }\wedge H_{\beta ]}\right)\nonumber\\
&=& {\cal \L\/}_n\,\vartheta ^\alpha\wedge\left(\,\Sigma ^{\rm matt}_\alpha +\Sigma ^{\rm em}_\alpha \right) +R_{\bot}^{\alpha\beta}\wedge \tau _{\alpha\beta}\,.\label{grender}
\end{eqnarray}
So, the energy exchange is performed in such a way that not the different types of energy separately, but only the sum (\ref{energycurr}) of all of them, (\ref{mattenergy}), (\ref{emenergy}) and (\ref{grenergy}), is conserved. The reason for it is that, in virtue of (\ref{covfieldeq2}), the total energy (\ref{energycurr}), although composed of three highly nontrivial pieces, reduces to an exact form as $\epsilon =-d\left(\vartheta _{\bot}^\alpha H_\alpha\right) +\vartheta _{\bot}^\alpha \left( DH_\alpha -\Pi _\alpha \right)\,$, which in general, contrarily to ${\cal H}$, is different from zero.

\section{Explicit Lagrangian pieces}

All the previous results were derived by invoking only the least action variational principle together with a symmetry principle. That is, until now we took into account two of the principles of Section II B, but we didn't miss the lacking principle expected to provide the form of the fundamental Lagrangian. However, in order to physically complete the formal scheme deduced previously, we finally have to introduce explicit gauge invariant Lagrangian pieces (\ref{Lagrangedecomp}), built from the covariant objects (\ref{tetraddef})--(\ref{covder2}), in order to derive the form of energy-momentum (\ref{definitions1(a)}) and of the generalized excitations (\ref{definitions1(b)})--(\ref{definition2}), as much as of concrete matter equations. In particular, for Dirac matter and for Maxwell electromagnetism we will use the corresponding standard Lagrangians, and for gravity a generalization of the Hilbert-Einstein one.

\subsection{Dirac matter}

Let us introduce the Dirac Lagrangian
\begin{equation}
L^{\rm matt}={i\over 2}\,(\,\overline{\psi}\,\,^*\gamma\wedge
D\psi +\overline{D\psi}\wedge\,^*\gamma\psi\,)
+\,^*m\,\overline{\psi}\psi\,,\label{Diraclagrang1}
\end{equation}
see \cite{Mielke:1997cb}, built with the Poincar\'e $\otimes$ $U(1)$ covariant derivatives (\ref{covder1}) and (\ref{covder2}), using the notation $\gamma :=\vartheta ^\alpha\,\gamma _\alpha\,$, with $\gamma ^\alpha$ as the Dirac gamma matrices, so that $\,{}^*\gamma :=\eta ^\alpha\,\gamma _\alpha \,$; see (\ref{antisym3form}). For a discussion on the absence of intrinsic translational contributions in such derivatives, see \cite{Tiemblo:2005sx}. From (\ref{Diraclagrang1}) we find
\begin{equation}
{{\partial L}\over{\partial d\overline{\psi}}}={i\over
2}\,{}^*\gamma\,\psi\,,\qquad {{\partial L}\over{\partial
d\psi}}={i\over 2}\,\overline{\psi}\,\,{}^*\gamma
\,,\label{signconventions}
\end{equation}
reflecting our sign conventions. The matter field equations take the form\footnote{With covariant derivatives defined as
$$D\left( {{\partial L}\over{\partial d\overline{\psi}}}\right):=
d\left( {{\partial L}\over{\partial d\overline{\psi}}}\right)
+i\,\left(\,eA-\Gamma ^{\alpha\beta}\sigma
_{\alpha\beta}\right)\wedge\left( {{\partial L}\over{\partial
d\overline{\psi}}}\right)\,,$$
$$D\left( {{\partial L}\over{\partial d\psi}}\right):= d\left(
{{\partial L}\over{\partial d\psi}}\right)+ \left( {{\partial
L}\over{\partial d\psi}}\right)\wedge i\,\left(\,eA-\Gamma
^{\alpha\beta}\sigma _{\alpha\beta}\right)\,.$$
In a more familiar notation, (\ref{mattfieldeq1}) and (\ref{mattfieldeq2}) read $$i\,{}^*\gamma\wedge D\,\psi -{i\over 2}\,D\eta ^\alpha\gamma _\alpha\psi +\,^*m\,\psi
=0\,,$$
$$i\,\overline{D\psi}\wedge\,{}^*\gamma +{i\over 2}\,\overline{\psi}\,D\eta ^\alpha\gamma _\alpha +\,^*m\,\overline{\psi}=0\,,$$
where
$$D\eta ^\alpha = \eta ^\alpha{}_\beta\wedge T^\beta\,,$$
see (\ref{antisym2form}) and (\ref{antisym3form}) with (\ref{torsiondef}).}
\begin{eqnarray}
D\left( {{\partial L}\over{\partial d\overline{\psi}}}\right)
-{i\over 2}\,{}^*\gamma\wedge D\,\psi -\,^*m\,\psi
&=&0\,,\label{mattfieldeq1}\\
D\left( {{\partial L}\over{\partial d\psi}}\right) + {i\over
2}\,\overline{D\psi}\wedge\,{}^*\gamma +\,^*m\,\overline{\psi}
&=&0\,.\label{mattfieldeq2}
\end{eqnarray}
By replacing (\ref{signconventions}) in (\ref{electcurr}) and (\ref{spincurr}) respectively, we get the explicit electric current
\begin{equation}
J =e\,\,\overline{\psi}\,\,^*\gamma\,\psi\,,\label{explicitelectcurr}
\end{equation}
satisfying (\ref{elcurrcons}), and the spin current
\begin{equation}
\tau _{\alpha\beta}=-{1\over 2}\,\overline{\psi}\,\left(\,
\sigma _{\alpha\beta}\,{}^*\gamma +\,{}^*\gamma\,{}\sigma _{\alpha\beta}\,\right)\,\psi \,,\label{explicitspincurr}
\end{equation}
entering (\ref{spincurrconserv}). Concerning the third matter current, namely matter energy-momentum, the following comment is in order. From the results previous to the introduction of the explicit Lagrangian, it is possible to derive two different expressions for the transversal part (\ref{trans-part}) of the matter energy-momentum. Indeed, from (\ref{spincurrconserv}) we find it to be
\begin{equation}
\underline{\Sigma}^{\rm matt}_{\alpha} = \vartheta _{\bot \alpha}\, \underline{\epsilon}^{\rm matt} -\underline{\vartheta}_\alpha\wedge \epsilon ^{\rm matt}_{\bot} -2\, \vartheta ^\beta _{\bot}\left( D\,\tau _{\alpha\beta}\,\right) _{\bot}
\,,\label{transversmattenmom1}
\end{equation}
resembling a phenomenological expression \cite{Obukhov:1993pt}, while (\ref{sigmamatt}) yields
\begin{equation}
\underline{\Sigma}^{\rm matt}_{\alpha} = \vartheta _{\bot \alpha}\, \underline{\epsilon}^{\rm matt} -\left( e_\alpha\rfloor\overline{\underline{D}\psi}\,\right){{\partial
L_{\bot}}\over{\partial {\it{l}}_n\overline{\psi}}} +{{\partial L_{\bot}}\over{\partial {\it{l}}_n\psi}}\left( e_\alpha\rfloor \underline{D}\,\psi\right)
\,.\label{transversmattenmom2}
\end{equation}
In both equations we made use of the decomposition
\begin{equation}
\epsilon ^{\rm matt}= N d\tau\wedge\epsilon ^{\rm matt}_{\bot} + \underline{\epsilon}^{\rm matt}\,,\label{energycurrfol}
\end{equation}
see (\ref{foliat1}), into the longitudinal and transversal parts of the matter energy
\begin{equation}
\epsilon ^{\rm matt} =\overline{{\cal \L\/}_n\psi}\,\,{{\partial
L}\over{\partial d\overline{\psi}}} -{{\partial L}\over{\partial
d\psi}}\,\,{\cal \L\/}_n\psi -L^{\rm matt}_{\bot}\,,\label{expmattenergy}
\end{equation}
found from (\ref{mattenergy}) with (\ref{sigmamatt}). The necessary coincidence between (\ref{transversmattenmom1}) and (\ref{transversmattenmom2}) in principle is not obvious, having to be imposed as a consistence condition. However, it follows automatically from (\ref{Diraclagrang1}). Actually, (\ref{sigmamatt}) results to be
\begin{equation}
\Sigma ^{\rm matt}_\alpha = (\,e_\alpha\rfloor {{\partial L}\over{\partial d\psi}}\,)\wedge D\psi -\overline{D\psi}\wedge (\,e_\alpha\rfloor {{\partial L}\over{\partial d\overline{\psi}}}\,) + m\,\eta _\alpha\overline{\psi}\psi
\,,\label{mattenergymom}
\end{equation}
for which both (\ref{transversmattenmom1}) and (\ref{transversmattenmom2}) hold simultaneously. This is due to the fact that, being the action explicitly gauge invariant, the consistence of all equations derived from it is guaranteed from the beginning.

Making use of (\ref{mattfieldeq1}) and (\ref{mattfieldeq2}) together with (\ref{mattenergymom}), we realize that $\vartheta ^\alpha\wedge\Sigma ^{\rm matt}_{\alpha} = \,^*m\,\overline{\psi}\psi\,$, and $L^{\rm matt} = 0\,$, so that in this case we also have  trivially ${\it l}_n\,L^{\rm matt} = 0\,$, implying, according to (\ref{L-Lie-der}), that the Dirac matter action does not evolve (horizontally), being time invariant.

\subsection{Electromagnetism}

Before introducing the invariant Maxwell Lagrangian, let us start the discussion of the electromagnetic case by deriving equations similar to (\ref{transversmattenmom1}) and (\ref{transversmattenmom2}) respectively. So, from (\ref{Symem-emt}) we deduce
\begin{equation}
\underline{\Sigma}^{\rm em}_{\alpha} = \vartheta _{\bot \alpha}\, \underline{\epsilon}^{\rm em} -\underline{\vartheta}_\alpha\wedge \epsilon ^{\rm em}_{\bot} \,,\label{transversemenmom1}
\end{equation}
while from (\ref{sigmaem}) we get
\begin{equation}
\underline{\Sigma}^{\rm em}_{\alpha} = \vartheta _{\bot \alpha}\, \underline{\epsilon}^{\rm em} +(\,e_\alpha\rfloor\underline{F}\,)\wedge\underline{H}\,,\label{transversemenmom2}
\end{equation}
involving the longitudinal and transversal components, analogous to those in (\ref{energycurrfol}), of the electromagnetic energy current
\begin{equation}
\epsilon ^{\rm em} = -F_{\bot}\wedge H -L^{\rm em}_{\bot}\,,\label{expemenergy}
\end{equation}
derived from (\ref{emenergy}) with (\ref{sigmaem}). We are interested in comparing (\ref{transversemenmom1}) and (\ref{transversemenmom2}) due to the fact that the consistence between both equations (as alternative expressions of a unique translative quantity) seems to require a relation of the Maxwell-Lorentz type between the electromagnetic excitation and the field strength. Indeed, the last term in (\ref{transversemenmom1}) reads explicitly
\begin{equation}
-\underline{\vartheta}_\alpha\wedge \epsilon ^{\rm em}_{\bot}
=-\underline{\vartheta}_\alpha\wedge F_{\bot}\wedge H_{\bot}
\,.\label{compar1}
\end{equation}
while (\ref{transversemenmom2}) can be rewritten with the help of (\ref{doublestar2}), (\ref{3dHodge2}) and (\ref{crossinterch2}), as
\begin{equation}
(\,e_\alpha\rfloor\underline{F}\,)\wedge\underline{H} =\underline{\vartheta}_\alpha\wedge{}^\#\underline{H}\wedge{}^\#\underline{F}
\,.\label{compar2}
\end{equation}
Comparison of (\ref{compar1}) and (\ref{compar2}) keeping in mind the decomposition (\ref{foliat2}) strongly suggests the proportionality $H\sim {}^*F$, not only as a sufficient condition, but even as a necessary consistence requirement. The Maxwell electromagnetic Lagrangian
\begin{equation}
L^{\rm em}=-{1\over 2}\,F\wedge\,^*F\,,\label{emlagrang1}
\end{equation}
see \cite{Hehl-and-Obukhov}, actually yields for the electromagnetic excitation (\ref{definitions1(b)}) the explicit form
\begin{equation}
H=-{{\partial L}\over{\partial dA}}={}^*F\,,\label{emmom}
\end{equation}
constituting the Maxwell-Lorentz electromagnetic spacetime relations \cite{Hehl-and-Obukhov}. It is by replacing (\ref{emmom}) in (\ref{covfieldeq1}) that we get the fundamental Maxwell equations.

From (\ref{emmom}) with (\ref{foliat2}) follows $H_{\bot} = {}^\#\underline{F}\,$, and $\underline{H} =-{}^\#F_{\bot}\,$, so that (\ref{transversemenmom1}) and (\ref{transversemenmom2}) become actually unified as
\begin{equation}
\underline{\Sigma}^{\rm em}_{\alpha} = \vartheta _{\bot \alpha}\, \underline{\epsilon}^{\rm em} -\underline{\vartheta}_\alpha\wedge F_{\bot}\wedge{}^\#\underline{F} \,.\label{transversemenmom3}
\end{equation}
On the other hand, the electromagnetic part (\ref{sigmaem}) of the momentum derived from the explicit Lagrangian (\ref{emlagrang1}) reads
\begin{equation}
\Sigma ^{\rm em}_\alpha = {1\over 2}\,\left[\,\left( e_\alpha\rfloor F\right)\wedge H -F\wedge\left( e_\alpha\rfloor H\right)\,\right]\,,\label{emenergymom}
\end{equation}
so that (\ref{expemenergy}) becomes finally
\begin{equation}
\epsilon ^{\rm em}=-\vartheta ^\alpha_{\bot}\Sigma ^{\rm em}_\alpha = -{1\over 2}\,\left[\,F_{\bot}\wedge H -F\wedge H_{\bot}\,\right]\,.\label{explemen1}
\end{equation}
In order to compare this result with more familiar notations \cite{Hehl-and-Obukhov}, we find the components of the electromagnetic energy current 3-form (\ref{explemen1}) analogous to the ones in (\ref{energycurrfol}) to be the energy flux or Poynting 2-form
\begin{equation}
\epsilon ^{\rm em}_{\bot}= F_{\bot}\wedge {}^\#\underline{F}\,,\label{explemen2}
\end{equation}
being identifiable as the exterior calculus version of the standard Poynting vector ${\buildrel {\rightarrow}\over E}\times {\buildrel {\rightarrow}\over B}\,$, and the energy density 3-form
\begin{equation}
\underline{\epsilon}^{\rm em}={1\over 2}\,\left( F_{\bot}\wedge {}^\#F_{\bot}
+\underline{F}\wedge {}^\#\underline{F}\,\right)\,,\label{explemen3}
\end{equation}
equal to the electromagnetic field energy which in standard vector notation reads ${1\over 2}\,(\,E^2 + B^2\,)\,dV$ . The {\it conservation equation} (\ref{emender}) can then be brought to a more explicit form by decomposing on the one hand
\begin{equation}
d\,\epsilon ^{\rm em}= N d\tau\wedge\left[\,{\it l}_n\,\underline{\epsilon}^{\rm em} -{1\over N}\,\underline{d}\left( N\epsilon ^{\rm em}_{\bot}\,\right)\,\right]\,,\label{energycurrder}
\end{equation}
using (\ref{longitderiv}), and on the other hand the remaining terms according to (\ref{foliat1}), to get
\begin{equation}
{\it l}_n\,\underline{\epsilon}^{\rm em} -{1\over N}\,\underline{d}\left( N\epsilon ^{\rm em}_{\bot}\,\right) = - {\cal \L\/}_n\,\vartheta ^\alpha _{\bot}\,\,\underline{\Sigma}^{\rm em}_{\alpha} -F_{\bot}\wedge J_{\bot}\,.\label{foliatemender1}
\end{equation}
Replacing (\ref{transversemenmom1}), and invoking the formal relation $N d\tau =-\vartheta _\alpha \vartheta ^\alpha _{\bot}\,$ found at the end of Appendix E to deduce $\underline{d}N/N =\vartheta ^\alpha _{\bot} (\,T_{\bot\alpha} - {\cal \L\/}_n\,\underline{\vartheta}^\alpha\,)\,$, we transform (\ref{foliatemender1}) into
\begin{equation}
{\it l}_n\,\underline{\epsilon}^{\rm em} = \underline{d}\,\epsilon ^{\rm em}_{\bot} +\vartheta ^\alpha _{\bot} (\,T_{\bot\alpha} - 2\,{\cal \L\/}_n\,\underline{\vartheta}^\alpha\,)\wedge\epsilon ^{\rm em}_{\bot} -F_{\bot}\wedge J_{\bot}\,.\label{foliatemender2}
\end{equation}
The time derivative of the energy density equals the divergence of the Poynting 2-form, plus additional terms having to do with the underlying geometry, plus a term which in standard notation reads ${\buildrel {\rightarrow}\over E}\cdot {\buildrel {\rightarrow}\over j}\,dV$, being interpretable as Joule's heat produced by the electric current \cite{Reitz-Milford}.

In parallel to the matter case, from (\ref{emenergymom}) we find $\vartheta ^\alpha\wedge\Sigma ^{\rm em}_{\alpha} = 0\,$, while $L^{\rm em} =-{1\over 2}\,F\wedge H\,$, being in principle ${\it l}_n\, L^{\rm em}\neq 0\,$, so that the electromagnetic action, according to (\ref{L-Lie-der}), evolves in time. (Actually, ${\it l}_n\, L^{\rm em} = d\,\epsilon ^{\rm em} -{1\over 2}\,\bigl( F\wedge J_{\bot} +F_{\bot}\wedge J\,\bigr)$.)

\subsection{Gravitation}

Since no universally accepted action exists for gravity, we take from Ref. \cite{Obukhov:2006ge} a quite general Lagrangian density including, besides a term of the Hilbert-Einstein type and a cosmological term, additional contributions quadratic in the Lorentz--irreducible pieces of torsion and curvature as established by McCrea \cite{Hehl:1995ue} \cite{McCrea:1992wa}. The gravitational Lagrangian reads
\begin{eqnarray}
L^{\rm gr}&=&{1\over{\kappa}}\,\left(\,\,{a_0\over
2}\,\,R^{\alpha\beta}\wedge\eta_{\alpha\beta}
-\Lambda\,\eta\,\right)\nonumber\\
&&-{1\over 2}\,\,T^\alpha\wedge
\left(\sum_{I=1}^{3}{{a_{I}}\over{\kappa}}\,\,{}^{*(I)}
T_\alpha\right)\nonumber\\
&&-{1\over 2}\,\,R^{\alpha\beta}\wedge\left(\sum_{I=1}^{6}b_{I}\,\,
{}^{*(I)}R_{\alpha\beta}\right)\,,\label{gravlagr}
\end{eqnarray}
with $\kappa$ as the gravitational constant, and $a_0$, $a_{I}$, $b_{I}$ as dimensionless constants. Definitions (\ref{antisym1form})--(\ref{eta4form}) are used, and the quadratic expressions are written taking into account that McCrea's irreducible torsion pieces ${}^{(I)}T_\alpha$ are mutually orthogonal, so that ${}^{(I)}T^\alpha\wedge\,{}^{*(I)}T_\alpha = T^\alpha\wedge\,{}^{*(I)}T_\alpha\,$. The same holds for the irreducible curvature pieces ${}^{(I)}R_{\alpha\beta}$. From (\ref{gravlagr}) we calculate the translational and Lorentz excitations (\ref{definitions1(c)}) and (\ref{definition2}) respectively to be
\begin{eqnarray}
H_\alpha &=& \sum_{I=1}^{3}{{a_{I}}\over{\kappa}}\,\,{}^{*(I)}
T_\alpha\,,\label{torsmom}\\
H_{\alpha\beta}&=&-{a_0\over{2\kappa}}\,\eta_{\alpha\beta} +\sum_{I=1}^{6}b_{I}\,\,
{}^{*(I)}R_{\alpha\beta}\,,\label{curvmom}
\end{eqnarray}
and we find the pure gravitational contribution (\ref{ealpha}) to the energy momentum
\begin{eqnarray}
E_\alpha &=& {a_0\over {4\kappa}}\,e_\alpha\rfloor \left(\,R^{\beta\gamma}\wedge\eta_{\beta\gamma}\,\right)-{\Lambda\over{\kappa}}\,\eta _\alpha\nonumber\\
&&+{1\over 2}\,\left[\,\left( e_\alpha\rfloor T^\beta\right)\wedge H_\beta  -T^\beta\wedge\left( e_\alpha\rfloor H_\beta \right)\,\right]\nonumber\\
&&+{1\over 2}\,\left[\,\left( e_\alpha\rfloor R^{\beta\gamma}\right)\wedge H_{\beta\gamma}  -R^{\beta\gamma}\wedge\left( e_\alpha\rfloor H_{\beta\gamma}\right)\,\right]\,.\nonumber\\
\label{gravenergymom}
\end{eqnarray}
(Notice the resemblance between (\ref{gravenergymom}) and (\ref{emenergymom}).) For completeness, let us also calculate the formulas analogous to (\ref{transversemenmom2}) and (\ref{expemenergy}) respectively. From (\ref{ealpha}) we get
\begin{eqnarray}
\underline{E}_{\alpha} &=& \vartheta _{\bot \alpha}\,\left(\,\underline{\epsilon}^{\rm gr} +\underline{D}\vartheta ^\beta _{\bot}\wedge\underline{H}_\beta\right)\nonumber\\
&&+(\,e_\alpha\rfloor\underline{T}^\beta\,)\wedge\underline{H}_\beta +(\,e_\alpha\rfloor\underline{R}^{\beta\gamma}\,)\wedge\underline{H}_{\beta\gamma}
\,,\label{transversgrenmom}
\end{eqnarray}
while (\ref{grenergy}) with (\ref{ealpha}) takes the form
\begin{equation}
\epsilon ^{\rm gr} = -{\cal \L\/}_n\vartheta ^\alpha \wedge H_\alpha -R_{\bot}^{\alpha\beta}\wedge H_{\alpha\beta} -L^{\rm gr}_{\bot}\,.\label{expgrenergy}
\end{equation}
On the other hand, the gravitational Lagrangian reduces to
\begin{equation}
L^{\rm gr} ={1\over 4}\,\vartheta ^\alpha\wedge E_\alpha -{1\over 2}\,T^\alpha\wedge H_\alpha -{1\over 2}\,R^{\alpha\beta}\wedge H_{\alpha\beta}\,,
\label{feqsealphaLag}
\end{equation}
with $\vartheta ^\alpha\wedge E_\alpha = {1\over{\kappa}}\,\left(\,a_0\,R^{\alpha\beta}\wedge\eta_{\alpha\beta}
-4\,\Lambda\,\eta\,\right)\,$.

For readers which are not familiar with exterior calculus notation \cite{Hehl:1995ue}, it may be useful to show how ordinary general relativistic Einstein equations are comprised as a particular case of the gauge-theoretical equations (\ref{covfieldeq2}) and (\ref{covfieldeq3}) with (\ref{torsmom}), (\ref{curvmom}) and (\ref{gravenergymom}). The Hilbert-Einstein theory in vacuum with cosmological constant derives from the pure gravitational Lagrangian (\ref{gravlagr}) with the constants fixed as $a_0=1\,$, $a_{I}=0\,$, $b_{I}=0\,$. Accordingly, (\ref{torsmom}) vanishes, (\ref{curvmom}) reduces to $H_{\alpha\beta}=-{1\over{2\kappa}}\,\eta_{\alpha\beta}\,$, and (\ref{gravenergymom}), coinciding with the whole energy-momentum (\ref{momentdecomp}) due to the absence of matter and radiation, becomes
\begin{equation}
E_\alpha = {1\over{\kappa}}\,\left(\,\,{1\over 2}\,\,R^{\beta\gamma}\wedge\eta_{\beta\gamma\alpha}
-\Lambda\,\eta _\alpha\,\right)\,.\label{H-Egravenergymom}
\end{equation}
The field equations (\ref{covfieldeq3}) then read
\begin{equation}
0=DH_{\alpha\beta} =-{1\over{2\kappa}}\, D\eta_{\alpha\beta} =-{1\over{2\kappa}}\,\eta_{\alpha\beta\gamma}\wedge T^\gamma\,,\label{H-Ecovfieldeq3}
\end{equation}
implying vanishing torsion, so that equations (\ref{covfieldeq2}) reduce to
\begin{eqnarray}
0&=&\Pi _\alpha =E_\alpha ={1\over{\kappa}}\,\left(\,\,{1\over 2}\,\,R^{\beta\gamma}\wedge\eta_{\beta\gamma\alpha}
-\Lambda\,\eta _\alpha\,\right)\nonumber\\
&=&-{1\over{\kappa}}\,e^i{}_\alpha\,\Bigl(\,
R_{\,ij}-{1\over 2}\, g_{\,ij}\,R +\Lambda\,g_{\,ij}\,\Bigr)\eta
^j\,,\label{H-Ecovfieldeq2}
\end{eqnarray}
constituting a well known reformulation of the ordinary Einstein equations in vacuum, which for clarity we also give in their standard form. For more details see for instance \cite{Tiemblo:2005js}.

\section{Hamiltonian approach to dynamics}

In Section IV we discussed covariant field equations as conditions derived from two complementary ways of imposing vertical invariance of the action, namely the principle of extremal action and the symmetry principle. In our exterior calculus notation, the coordinate independent field equations (\ref{covfieldeq1})--(\ref{covfieldeq3}) do not display any explicit reference to the base space. But in Section VI we introduced a base space foliation becoming actually reflected in the notation (even in the language of differential forms) by distinguishing from each other the projections respectively longitudinal and transversal with respect to a certain {\it parametric time} direction (defined in the base space).

Associated with such foliation, we presented {\it parametric time evolution} as a form of horizontal displacement on the base space, compatible with the field equations guaranteeing vertical invariance. So to say, vertical invariance guides horizontal motions. Bundle connections (that is, gauge potentials) are known to define horizontality in fiber bundles. Thus, provided the field equations hold, connections become responsible for maintaining several vertical features along horizontal paths in the base space. Vertical invariance conditions act as {\it forces} or {\it interactions} influencing the quantities subjected to horizontal evolution displacements. Here we briefly outline a Hamiltonian formalism suitable to deal with evolution understood in this manner.

\subsection{The Hamiltonian evolution equations}

In the present approach, a central role is played by the vanishing Hamiltonian-like 3-form (\ref{EC06}), whose transversal part reads
\begin{equation}
\underline{\cal H} =\,Q_{\bot}\wedge{{\partial
L_{\bot}}\over{\partial Q_{\bot}}} + \left(
d\,Q\right)_{\bot}\wedge {{\partial L_{\bot}}\over{\partial
{\it{l}}_n\underline{Q}}}
-L_{\bot}-Q_{\bot}\wedge {{\delta L_{\bot}}\over{\delta Q_{\bot}}}
\,,\label{transversH}
\end{equation}
being covariant as a consequence of the symmetry conditions (\ref{consistencecond1})--(\ref{consistencecond3}) and (\ref{consistencecond4}); compare with the non-foliated expression (\ref{EC12}). The relevance of the quantity (\ref{transversH}) derives from the fact that it results to occur in the variational formula (\ref{varlag1}) when foliated as (\ref{varlag3}), so that by taking into account (\ref{longitderiv}) as much as the foliated field equations (\ref{foliatfieldeq1}) and (\ref{foliatfieldeq2}), (\ref{varlag3}) yields
\begin{eqnarray}
0&=&N d\tau\wedge \Bigl[\,\delta\underline{\cal H} +\delta\underline{Q}\wedge
{\it{l}}_n\left(\,{{\partial L_{\bot}}\over{\partial {\it{l}}_n\underline{Q}}}\,\right)
-{\it{l}}_n\underline{Q}\wedge\delta\left(\,{{\partial L_{\bot}}\over{\partial {\it{l}}_n\underline{Q}}}\,\right)\nonumber\\
&&\hskip1.8cm +\delta Q_{\bot}\wedge {{\delta
L_{\bot}}\over{\delta Q_{\bot}}} + \delta\underline{Q}\wedge
{{\delta L_{\bot}}\over{\delta\underline{Q}}}\,\Bigr]\nonumber\\
&& -d \left\{\,N d\tau\wedge\left[\,
Q_{\bot}\wedge\delta\left(\,{{\partial L_{\bot}}\over{\partial {\it{l}}_n\underline{Q}}}\,\right) +\delta\underline{Q}\wedge {{\partial
L_{\bot}}\over{\partial\underline{dQ}}}\,\right]\,\right\}\,.\nonumber\\
\label{varlag8}
\end{eqnarray}
Now we introduce the momentum notation
\begin{equation}
\,{}^{\#}\pi ^{_{\underline{Q}}}:= {{\partial
L_{\bot}}\over{\partial {\it{l}}_n\underline{Q}}}\,,\label{qmom}
\end{equation}
following previous work \cite{Tiemblo:2005js} \cite{Lopez-Pinto:1997aw} \cite{Wallnerphd}\footnote{We conserve in our notation the Hodge dual star ${\#}$ in three dimensions in order to facilitate comparison with the literature, although this detail may be irrelevant in the present context.}. By imposing the divergence term in (\ref{varlag8}) to vanish at the boundary in analogy to the divergence term in (\ref{varlag4}), and provided the field equations hold, from (\ref{varlag8}) with (\ref{qmom}) we read out
\begin{equation}
\delta \underline{\cal H} = -\delta\underline{Q}\wedge
{\it{l}}_n\,{}^{\#}\pi ^{_{\underline{Q}}} +
{\it{l}}_n\underline{Q}\wedge\delta\,{}^{\#}\pi
^{_{\underline{Q}}}\,.\label{Hamiltvar1}
\end{equation}
Next we take the Hamiltonian 3-form to be a functional $\underline{\cal H}= \underline{\cal H}\left(\,Q_{\bot}\,,\underline{Q}\,\,;\,{}^{\#}\pi ^{_{Q_{\bot}}}\,,{}^{\#}\pi ^{_{\underline{Q}}}\,\,\right)\,$, so that by applying the chain rule \cite{Lopez-Pinto:1997aw} \cite{Wallnerphd} we get
\begin{eqnarray}
\delta \underline{\cal H}&=&\delta Q_{\bot}\wedge {{\partial \underline{\cal
H}}\over{\partial Q_{\bot}}} +\delta\underline{Q}\wedge {{\partial \underline{\cal
H}}\over{\partial\underline{Q}}}\nonumber\\
&+&{{\partial \underline{\cal H}}\over{\partial
\,{}^{\#}\pi ^{_{Q_{\bot}}}}} \wedge\delta\,{}^{\#}\pi
^{_{Q_{\bot}}}\,\,+ {{\partial \underline{\cal H}}\over{\partial
\,{}^{\#}\pi ^{_{\underline{Q}}}}} \wedge\delta\,{}^{\#}\pi
^{_{\underline{Q}}}\,\,.\label{Hamiltvar2}
\end{eqnarray}
By comparing (\ref{Hamiltvar1}) with (\ref{Hamiltvar2}) we find the Hamiltonian evolution equations
\begin{eqnarray}
0&=&{{\partial \underline{\cal H}}\over{\partial Q_{\bot}}}\,,\qquad
{\it{l}}_n\,{}^{\#}\pi ^{_{\underline{Q}}} =
-{{\partial \underline{\cal H}}\over{\partial\underline{Q}}}\,,\label{Hamilteq1}\\
0&=&{{\partial \underline{\cal H}}\over{\partial \,{}^{\#}\pi
^{_{Q_{\bot}}}}}\,,\qquad {\it{l}}_n\underline{Q} = {{\partial \underline{\cal
H}}\over{\partial \,{}^{\#}\pi ^{_{\underline{Q}}}}}
\,.\label{Hamilteq2}
\end{eqnarray}
(Left equations in (\ref{Hamilteq1}) and (\ref{Hamilteq2}) are to be interpreted respectively as ${\it{l}}_n\,{}^{\#}\pi ^{_{Q_{\bot}}} =0$ and ${\it{l}}_n\,Q_{\bot} =0$.) Eqs. (\ref{Hamilteq1}), (\ref{Hamilteq2}) describe Hamiltonian {\it parametric time evolution} of any dynamical quantity as given by its Lie derivative along the time-like vector field $n$. Evolution is generated by the vanishing quantity (\ref{transversH}) --we recall that it is the transversal part of (\ref{EC12}) and thus of (\ref{vanish-H})--, which reveals to play the role of an evolution operator.

Generalized Poisson brackets can be introduced \cite{Tiemblo:2005js} \cite{Lopez-Pinto:1997aw} \cite{Wallnerphd}, applicable to arbitrary dynamical quantities represented by differential forms. Denoting by $\Phi$ either longitudinal or transversal components (\ref{long-part}), (\ref{trans-part}) of $p$-forms, their evolution is given by
\begin{eqnarray}
{\it{l}}_n\,\Phi
=\,\left\{\,\Phi\,,\underline{\cal{H}}\,\right\} &:=& {{\partial
\underline{\cal{H}}}\over{\partial\,{}^\#\pi ^{^{Q_{\bot}}}}}\wedge {{\partial
\Phi}\over{\partial Q_{\bot}}} -{{\partial\Phi}\over{\partial\,{}^\#\pi ^{^{Q_{\bot}}}}}\wedge {{\partial\underline{\cal{H}}}\over{\partial Q_{\bot}}}\nonumber \\
&&+{{\partial \underline{\cal{H}}}\over{\partial\, {}^\#\pi
^{^{\underline{Q}}}}}\wedge {{\partial \Phi}\over{\partial
\underline{Q}}} -{{\partial \Phi}\over{\partial\, {}^\#\pi
^{^{\underline{Q}}}}} \wedge {{\partial \underline{\cal{H}}}\over{\partial
\underline{Q}}}\,.\nonumber\\
\label{Poisson1}
\end{eqnarray}
More rigorously one should define Poisson brackets for differential forms as
\begin{eqnarray}
\bigl\{\alpha\bigl( x\,\bigr),\, \beta\bigl( y\,\bigr)\bigr\}
&:=&\int _z \Bigl[{{\partial\,\beta\bigl( y\,\bigr)}
\over{\partial\,{}^\#\pi _i\bigl( z\,\bigr)}} \wedge
{{\partial\,\alpha\bigl( x\,\bigr)} \over{\partial Q^i\bigl(
z\,\bigr)}}\nonumber\\
&&-{{\partial\,\alpha\bigl( x\,\bigr)}
\over{\partial\,{}^\#\pi _i\bigl( z\,\bigr)}} \wedge
{{\partial\,\beta\bigl( y\,\bigr)} \over{\partial Q^i\bigl(
z\,\bigr)}}\,\Bigr] \wedge\overline{\eta}\bigl( z\,\bigr)
\,,\nonumber\\
\label{Poisson2}
\end{eqnarray}
see \cite{Wallnerphd}, with the arbitrary forms $\alpha $ and $\beta $ representing functionals of the canonical conjugate variables concisely denoted as $Q^i\,$, ${}^\#\pi _i\,$. From (\ref{Poisson2}) we find $\bigl\{ Q^i(x\,)\,,Q^j(y\,)\,\bigr\} =\,0\,$, $\bigl\{ {}^\#\pi _i(x\,)\,,{}^\#\pi _j(y\,)\,\bigr\} =\,0\,$, and $\bigl\{ Q^i(x\,)\,,{}^\#\pi _j(y\,)\,\bigr\} =\,\delta ^i_j\,\delta ^3(x-y\,)\,$, as expected.

{\it Parametric time} evolution as given by (\ref{Poisson1}) is evaluated along the time-like vector $n$, the latter constituting a non-dynamical object defined on the base space. Nevertheless, one can alternatively introduce {\it clock time} as a suitable dynamical quantity, in such a way that {\it clock time} evolution becomes expressed as a relation between fiber variables (\ref{constfields}). A quite natural choice of such an internal {\it time} is that of the component $\xi ^0$ of the coordinate-like fields $\xi ^\alpha$. Equation (\ref{Poisson1}) can be reformulated so that the Lie derivative ${\it{l}}_n$ becomes replaced by a derivative with respect to $\xi ^0$, see \cite{Tiemblo:2002uk}. The price one pays by doing so is that explicit covariance gets lost. A different physical time choice respecting covariance was presented in \cite{Lopez-Pinto:1997aw}.

\subsection{Covariance and symmetry generators}

Gauge theories are constrained systems \cite{Hanson:1976}. We are going to show briefly the form of the first class constraints acting as symmetry generators in Dirac's Hamiltonian approach \cite{Dirac50} in the particular case of the gauge theory of Poincar\'e $\otimes$ $U(1)$. Such constraints are generalized Gauss laws corresponding respectively to $U(1)\,$, to the Lorentz group, and to translations, the latter ones behaving in close analogy to the remaining symmetries. We aren't going to develop the full Hamiltonian formalism, but in order to outline it we have to use the momentum notation (\ref{qmom}) summarizing the various momenta
\begin{eqnarray}
\,{}^{\#}\pi ^{\xi}_\alpha &:=& {{\partial L_{\bot}}\over{\partial
{\it{l}}_n\xi ^\alpha}}\,,\quad
\,{}^{\#}\pi _{\overline{\psi}}:= {{\partial
L_{\bot}}\over{\partial
{\it{l}}_n\overline{\psi}}}\,,\quad
\,{}^{\#}\pi _{\psi}:= {{\partial L_{\bot}}\over{\partial
{\it{l}}_n\psi}}\,,\label{moms1}\\
\,{}^{\#}\pi ^{\underline{A}}&:=& {{\partial L_{\bot}}\over{\partial
{\it{l}}_n\underline{A}}}\,,\quad \,{}^{\#}\pi ^{\buildrel (T)\over{\underline\Gamma}}_\alpha:=
{{\partial L_{\bot}}\over{\partial {\it{l}}_n {\underline{\Gamma}_{(T)}^\alpha}}}\,,\quad
\,{}^{\#}\pi ^{\underline{\Gamma}}_{\alpha\beta}:= {{\partial
L_{\bot}}\over{\partial {\it{l}}_n\underline{\Gamma
}^{\alpha\beta}}}\,,\nonumber\\
\label{moms2}
\end{eqnarray}
as much as other possible ones such as $\,{}^{\#}\pi ^{A_{\bot}}:= {{\partial L_{\bot}}\over{\partial {\it{l}}_n A_{\bot}}}$ which in any gauge theory are automatically equal to zero. On the other hand, according to (\ref{condit2}) we decompose for instance (\ref{definitions1(b)}) as $H=-{{\partial L}\over{\partial dA}}=- ( N d\tau\wedge {{\partial L_{\bot}}\over{\partial\underline{dA}}} +{{\partial L_{\bot}}\over{\partial {\it{l}}_n\underline{A}}}\,)\,$, so that with the first definition in (\ref{moms2}) we find $\underline{H}=-\,{}^{\#}\pi ^{\underline{A}}$, while from (\ref{definitions1(c)}) and (\ref{definition2}) with (\ref{moms2}) we get $\underline{H}_\alpha =-\,{}^{\#}\pi ^{\buildrel (T)\over{\underline\Gamma}}_\alpha\,$ and $\underline{H}_{\alpha\beta}=-(\,{}^{\#}\pi ^{\underline{\Gamma}}_{\alpha\beta}-\xi _{[ \alpha} {}^{\#}\pi ^{\buildrel (T)\over{\underline\Gamma}}_{\beta ]}\,)\,$. By replacing (\ref{moms1})--(\ref{moms2}) as much as the symmetry conditions (\ref{consistencecond1})--(\ref{consistencecond3}) into the vanishing Hamiltonian 3-form (\ref{transversH}), the latter takes the Lorentz covariant form
\begin{eqnarray}
\underline{\cal H} =&&\vartheta _{\bot}^\alpha \,{}^{\#}\pi ^{\xi}_\alpha +\overline{{\cal
\L\/}_n\psi}\,\,\,{}^{\#}\pi _{\overline{\psi}}\,\,\, -\,\,{}^{\#}\pi _{\psi}\,\,{\cal \L\/}_n\psi\,\,+F_{\bot}\wedge {}^{\#}\pi ^{\underline{A}}\nonumber\\
&&+ T_{\bot}^\alpha\wedge\,{}^{\#}\pi ^{\buildrel
(T)\over{\underline\Gamma}}_\alpha\,\,+
R_{\bot}^{\alpha\beta}\wedge\Bigl(\,{}^{\#}\pi
^{\underline{\Gamma}}_{\alpha\beta}-\xi _{[\alpha}{}^{\#}\pi
^{\buildrel (T)\over{\underline\Gamma}}_{\beta ]}\,\Bigr) -L_{\bot}\nonumber\\
&&-\Bigl(\,A_{\bot}\,\hat{C} +{\buildrel (T)\over{\Gamma
_{\bot}^\alpha}}\,\hat{P}_\alpha +\Gamma
_{\bot}^{\alpha\beta}\,\hat{L}_{\alpha\beta}\,\Bigr)\,.\label{transversHexplicit}
\end{eqnarray}
The details of the dynamical approach based on (\ref{transversHexplicit}) will be developed elsewhere, constituting a modified version of the Hamiltonian formalism already published in Refs. \cite{Tiemblo:2005js} \cite{Lopez-Pinto:1997aw}, with the difference that the present formalism is adapted to a different explicit covariance.

The last terms in (\ref{transversHexplicit}) constitute the explicit expansion of the term $Q_{\bot}\wedge\bigl({\delta L_{\bot}}/{\delta Q_{\bot}}\bigr)$ in (\ref{transversH}), proportional to the transversal parts of the field equations. The latter ones vanish separately as much as the remaining {\it Hamiltonian} 3-form (\ref{transversH}) does. The main reason for keeping them in (\ref{transversHexplicit}) is that they play the role of first class constraints, and thus of generators of the symmetries involved in the theory \cite{Hanson:1976}. Since the $Q$'s given in (\ref{constfields}) are either 0-forms or 1-forms, the transversal parts $Q_{\bot}$ only exist as 0-forms. In particular, they are the longitudinal parts of the gauge potentials, that is $A_{\bot}$, ${\buildrel (T)\over{\Gamma _{\bot}^\alpha}}$ and $\Gamma _{\bot}^{\alpha\beta}\,$, playing the role of Lagrange multipliers. The constraints present in (\ref{transversHexplicit}) as the generators of Poincar\'e $\otimes$ $U(1)$ read
\begin{eqnarray}
\hat{C} &:=&\underline{d}\,{}^{\#}\pi ^{\underline{A}}
\,-ie\,\Bigl(\,\overline{\psi}\,\,{}^{\#}\pi _{\overline{\psi}}\,+
\,\,{}^{\#}\pi _{\psi}\,\psi\,\Bigr)
\,,\label{chargeconstr}\\
\hat{P}_\alpha &:=&\underline{D}\,{}^{\#}\pi ^{\buildrel
(T)\over{\underline\Gamma}}_\alpha \,+\,{}^{\#}\pi
^{\xi}_\alpha\,,\label{linmomconstr}\\
\hat{L}_{\alpha\beta} &:=&\underline{D}\,{}^{\#}\pi
^{\underline{\Gamma}}_{\alpha\beta} \,+{\buildrel
(T)\over{\underline\Gamma}}_{[\alpha}\wedge {}^{\#}\pi ^{\buildrel
(T)\over{\underline\Gamma}}_{\beta ]} \,+\xi\,_{[\alpha}
{}^{\#}\pi ^{\xi}_{\beta ]}\nonumber\\
&&\hskip0.3cm +i\,\Bigl(\,\overline{\psi}\,\sigma
_{\alpha\beta}\,{}^{\#}\pi _{\overline{\psi}}\,+\,\,{}^{\#}\pi
_{\psi}\,\sigma _{\alpha\beta}\,\psi\,\Bigr)
\,,\label{angmomconstr}
\end{eqnarray}
being identical with the transversal part of the field equations in their form (\ref{coveq1})--(\ref{coveq3}). The covariantized form of (\ref{angmomconstr}) is obtained by combining (\ref{linmomconstr}) and (\ref{angmomconstr}) into
\begin{eqnarray}
\hat{L}_{\alpha\beta}-\xi\,_{[\alpha}\,\hat{P}_{\beta
]} &=&\underline{D}\Bigl(\,{}^{\#}\pi
^{\underline{\Gamma}}_{\alpha\beta} -\xi _{[\alpha }{}^{\#}\pi
^{\buildrel (T)\over{\underline\Gamma}}_{\beta ]}\,\Bigr)
+\underline{\vartheta}_{[\alpha}\wedge {}^{\#}\pi ^{\buildrel
(T)\over{\underline\Gamma}}_{\beta ]}\nonumber\\
&&+ i\,\Bigl(\,\overline{\psi}\,\sigma _{\alpha\beta}\,{}^{\#}\pi
_{\overline{\psi}}\,+\,\,{}^{\#}\pi _{\psi}\,\sigma
_{\alpha\beta}\,\psi\,\Bigr)\,.\nonumber\\
\label{transversfieldeq}
\end{eqnarray}
(Compare with the transversal part of (\ref{covfieldeq3}).) By building a symmetry generator with the form of the last terms in (\ref{transversHexplicit}) with the $Q_{\bot}$'s replaced by the usual group parameters, that is
\begin{equation}
\hat{G}:=-\left(\,{\lambda\over e}\,\,\hat{C} +\epsilon ^\mu\,\hat{P}_\mu +\beta ^{\mu\nu}\,\hat{L}_{\mu\nu}\,\right)\,,\label{symmgenerator}
\end{equation}
variations of any dynamical variable can be obtained with the help of Poisson brackets (\ref{Poisson2}) as
\begin{equation}
\delta \alpha =\{\,\alpha\,,\hat{G}\}\,.\label{varalpha}
\end{equation}
In particular mainly due to the fact that --up to Dirac deltas--
\begin{equation}
\{\,\xi^\alpha\,,\hat{P}_\beta\}=\{\,\xi^\alpha\,,\,{}^{\#}\pi ^{\xi}_\beta\}=\delta ^\alpha _\beta\,,\label{xipi}
\end{equation}
we are able to reproduce (\ref{varcoordGoldstone}) as
\begin{equation}
\delta \xi ^\alpha =\{\,\xi ^\alpha\,,\hat{G}\} =-\,\xi ^{\,\beta}\beta _\beta {}^\alpha -\epsilon ^\alpha\,,\label{varxi}
\end{equation}
and analogously we can calculate the remaining variations (\ref{varpsi})--(\ref{varlorconn}).

\section{Final remarks}

At the end of Section VIII A, we mentioned an example of loss of explicit symmetry --without symmetry breaking-- associated with the choice of $\xi ^0$ as {\it clock time}. At this point, let us mention further cases of explicit symmetry loss which also result to be useful. For instance, one can find certain similitudes between the gauge equations introduced above and related equations of Classical Mechanics. We begin by reformulating (\ref{sigmamattconserv}) as a force law
\begin{equation}
D\,\Sigma ^{\rm matt}_\alpha = f_\alpha\,,\label{reformsigmamattconserv}
\end{equation}
(obtained by applying the symmetry principle separately to the matter Lagrangian) with $f_\alpha$ being understood as an external force 4-form generalizing the Lorentz force. Using (\ref{reformsigmamattconserv}) and (\ref{tetrad}), it is also possible to rewrite (\ref{spincurrconserv}) as an equation for generalized angular momentum
\begin{equation}
D \left(\,\tau _{\alpha\beta} +\xi _{[\,\alpha}\wedge\Sigma
^{\rm matt}_{\beta ]}\right) +{\buildrel (T)\over{\Gamma _{[ \,\alpha}}}\wedge\Sigma ^{\rm matt}_{\beta ]}= \xi _{[\,\alpha} f_{\beta ]}\,,\label{reformspincurrconserv}
\end{equation}
where the term in the r.h.s. behaves as a generalized torque.

Renouncing to explicit covariance also helps in finding strictly conserved currents from the covariant quasi-conservation equations (\ref{covfieldeq1})--(\ref{covfieldeq3}). Actually, true conservation as expressed by the continuity equations (\ref{elcurrcons}) and (\ref{energyconserv}) involve ordinary differentials rather than covariant ones, so that exact conservation of tensor quantities cannot be formulated covariantly. Thus let us reformulate the covariant equations (\ref{covfieldeq1})--(\ref{covfieldeq3}) in terms of suitable currents as follows. With the help of definitions (\ref{definitions1(a)})--(\ref{definition2}), we leave (\ref{consistencecond1}) as it is but from (\ref{consistencecond2}) and (\ref{consistencecond3}) we define respectively the noncovariant linear momentum current
\begin{equation}
J_\alpha := {{\partial L}\over{\partial \Gamma _{^{(\,T)}}^\alpha }} =\Pi _\alpha +\Gamma _\alpha{}^\beta\wedge H_\beta\,,\label{noncovlinmom}
\end{equation}
and the noncovariant angular momentum current
\begin{eqnarray}
J_{\alpha\beta} :={{\partial L}\over{\partial \Gamma ^{\alpha\beta}}} =&& \tau _{\alpha\beta} +\xi _{[\,\alpha } \Pi _{\beta ]}
-{\buildrel (T)\over{\Gamma _{[\,\alpha }}}\wedge H_{\beta ]}\nonumber\\
&&+\Gamma _\alpha{}^\gamma\wedge \left( H_{\gamma\beta} +\xi _{[ \gamma } H_{\beta ]}\right)\nonumber\\
&&-\Gamma _\beta{}^\gamma\wedge \left( H_{\gamma\alpha} +\xi _{[ \gamma } H_{\alpha]}\right)\,,\label{noncovangmom}
\end{eqnarray}
so that the covariant field eqs. (\ref{covfieldeq1})--(\ref{covfieldeq3}) become expressible as
\begin{eqnarray}
dH &=&J\,,\label{noncovfieldeq1}\\
dH_\alpha &=& J_\alpha\,,\label{noncovfieldeq2}\\
d\left( H_{\alpha\beta} +\xi _{[ \alpha } H_{\beta]}\right)
&=& J_{\alpha\beta} +\xi _{[\,\alpha }\left( dH_{\beta ]}-J_{\beta ]}\right)
\,.\label{noncovfieldeq3}
\end{eqnarray}
Obviously, from (\ref{noncovfieldeq1})--(\ref{noncovfieldeq3}) follow the true conservation equations
\begin{eqnarray}
dJ &=&0\,,\label{cons-1}\\
dJ_\alpha &=& 0\,,\label{cons-2}\\
d J_{\alpha\beta}&=& 0\,.\label{cons-3}
\end{eqnarray}
On the other hand, let us end this section mentioning the possible relevance of translations for interpreting the position-momentum commutation relations of Quantum Mechanics. Indeed, the analogy between (\ref{xipi}) and the commutation relations
\begin{equation}
\left[\,\Xi ^{\alpha}\,, P_{\beta}\,\right] =\,i\,\delta ^\alpha _\beta\,,\label{QM}
\end{equation}
might allow to regard (\ref{QM}) as the reformulation of a translational property concerning $\xi ^\alpha$ and $\hat{P}_\beta\,$ --or maybe ${}^{\#}\pi ^{\xi}_\beta$-- into the language of operators, with $\Xi ^\alpha$ as the operator version of our position vector $\xi ^\alpha$. Notice in fact that, by introducing $G_{\rm trans}:= i\,\epsilon ^\mu P_{\mu}$ as the generator of translations, similar to the corresponding piece in (\ref{symmgenerator}), we get
\begin{equation}
\delta \Xi ^{\alpha} = \left[\,\Xi ^{\alpha}\,, G_{\rm trans}\right] =-\epsilon ^\alpha\,,\label{QMtranslat}
\end{equation}
as a translational-like variation, while a Poincar\'e generator, say $G_{\rm Poinc}:= i\,\left(\,\epsilon ^\mu P_{\mu} +\beta ^{\mu\nu}L_{\mu\nu}\,\right)$, with $L_{\mu\nu}:= \Xi _{[\,\mu}P_{\nu ]}$, yields
\begin{equation}
\delta \Xi ^{\alpha} = \left[\,\Xi ^{\alpha}\,,G_{\rm Poinc}\right] =-\beta _\beta{}^\alpha\,\Xi ^\beta -\epsilon ^\alpha\,,\label{QMPoinc}
\end{equation}
analogous to the field variation (\ref{varxi}).

\section{Conclusions}

Translations are an usually forgotten symmetry in the context of gauge-theoretical dynamics of fundamental interactions. We have shown that, although hidden, they are present in a variety of physical contexts. So, in Newtonian Mechanics, global space translations are responsible for linear momentum conservation. Due to the fact that the same rigid displacement $\epsilon ^a$ in (\ref{CM3}) makes sense simultaneously at distant positions, a momentum interchange is predicted to occur between far separated bodies, thus providing a basis for action at a distance. In gauge theories instead, group parameters depend on base space coordinates. Local spacetime translations $\epsilon ^\alpha (x)\,$ are different at different points, so that only local interchanges --say "collisions"-- are admissible. Consequently, action at a distance abandons the scene in favor of an interchange of linear momentum affecting fields locally: interactions mediated by gauge potentials replace remote influence in Newton's manner.

Fiber bundles are known to be the geometrical structures underlying Yang-Mills theories of internal local groups \cite{Trautman:1970cy} \cite{Bleecker} \cite{Kobayashi} \cite{Wu:1975es} \cite{Daniel:1979ez} \cite{Eguchi:1980jx}. A slight modification of them also constitutes the implicit geometrical background of the present paper. Bundles merely have to be made enough flexible to accommodate local translations conveniently. Indeed, by embracing the translational group as a gauge symmetry, we accept it to be fully distinguished from horizontal (base space) diffeomorphisms, since gauge transformations are vertical. Notwithstanding, there is possible for translations to actively move from a spacetime position to another provided the affected points aren't presupposed to be identical with base space ones. (See Appendices B and C.) In Ref. \cite{Tresguerres:2002uh} we proposed a certain composite bundle as the geometrical framework suitable to deal with the local realization of translations. The fibers of composite bundles are to be visualized as broken lines, with the translational sector (attached itself to the base space) acting as an intermediate base space where other fiber sectors orthogonal to it are attached to. Translations become unified with any internal symmetry, so that all interactions including gravitation can be treated in a homogeneous gauge-theoretical way within a unique structure. As shown in the present paper, geometry and light --gravity and radiation-- appear as different aspects of the same unified bundle approach to spacetime and internal forces, all of them obeying similar field equations; see (\ref{covfieldeq1})--(\ref{covfieldeq3}).

The bundle is equipped with an action required to be vertically invariant. Horizontal displacements are subjected to interactions in the sense that they must respect the vertical invariance conditions, that is the (covariant) field equations. However, in our proposal the base space is not {\it dynamical}, but it plays the role of a sort of inert screen. Not a base space metric, but tetrads defined on the fibres, and in general quantities built from the fiber variables (\ref{constfields}), are the physical objects affected by dynamical laws. This makes a major difference with respect to ordinary General Relativity, where spacetime is modelized by a manifold equipped with a dynamical metric, being such dynamical spacetime expected to act as the base space of Yang-Mills theories of internal groups when gravity is present. Certainly, in our case as well as in General Relativity, dynamics manifests itself on the base space, where evolution occurs as a horizontal consequence of vertical invariance; also a Riemannian metric (dynamically determined), and thus a full Riemannian geometry, can be defined on our base space. But, remarkably, only as the result of the pullback of the vertical structures considered in the present paper. See \cite{Tresguerres:2002uh} for more details.

The coordinate-like translational Goldstone fields $\xi ^\alpha$ taken from the nonlinear Poincar\'e Gauge Theory \cite{Tiemblo:2005js} \cite{Tresguerres:2002uh} were shown to play a central role due to the nonminimal universal coupling of these translational variables --and thus of gravity-- to any other quantity. This fact mainly manifests itself in the contribution of all dynamical fields to the energy-momentum $\Pi _\alpha$ and accordingly to the energy current 3-form (\ref{energycurr}). (See also Appendix C for a discussion on the relevance of the fields $\xi ^\alpha$ for the description of motion.) Finally, we recall that we explained why translations, despite their fundamental contributions as made manifest in the present paper, remain a hidden and commonly ignored symmetry, as a consequence of the translation-invariant structure (\ref{tetrad}) of the tetrads.


\begin{acknowledgments}
The author wants to thank Alfredo Tiemblo for past collaboration.
\end{acknowledgments}


\appendix
\section{Definitions of derived dynamical quantities}

In the main text we made use of the following definitions. On the one hand
the combination
\begin{equation}
\vartheta ^\alpha := D\,\xi ^\alpha +{\buildrel (T)\over{\Gamma
^\alpha}} = d\,\xi ^\alpha + \Gamma
_\beta{}^\alpha\,\xi ^\beta +{\buildrel (T)\over{\Gamma
^\alpha}}\label{tetraddef}
\end{equation}
provides us with a modified translational gauge potential which
turns out to be invariant under translations, transforming
as a Lorentz covector, see (\ref{var-1}) below. In \cite{Julve:1994bh} \cite{Tresguerres:2000qn} \cite{Tiemblo:2005js} \cite{Tresguerres:2002uh} we discussed (\ref{tetraddef}) as the components of translational nonlinear connections, which we identified as tetrads. So we do here. Notice that in the absence of connections and thus of gravity, that is, in the Minkowski space, (\ref{tetraddef}) reduces to the trivial tetrad $d\,\xi ^\alpha$. Torsion is defined \cite{Hehl:1974cn} \cite{Hehl:1976kj} \cite{Hehl:1979xk} \cite{Hehl:1995ue} as the covariant differential of tetrads (\ref{tetraddef}), namely
\begin{equation}
T^\alpha := D\,\vartheta ^\alpha = d\,\vartheta ^\alpha + \Gamma _\beta{}^\alpha\wedge\vartheta
^\beta\,,\label{torsiondef}
\end{equation}
while the definition of the Lorentzian curvature reads
\begin{equation}
R_\alpha{}^\beta := d\,\Gamma _\alpha{}^\beta + \Gamma
_\gamma{}^\beta\wedge \Gamma _\alpha{}^\gamma\,,\label{curvdef}
\end{equation}
being antisymmetric in the indices $\alpha\,,\beta\,$. Such anholonomic Lorentz indices are rised and lowered with the help of the anholonomic constant Minkowski metric $o_{\alpha\beta}= diag(-+++)$ which is assumed to exist as the natural invariant of the local Poincar\'e group ($\delta o_{\alpha\beta} =0\,$).

Besides these quantities, we also define the ordinary electromagnetic field strength
\begin{equation}
F:= dA\,,\label{Fdef}
\end{equation}
and the covariant derivatives of matter fields
\begin{eqnarray}
D\psi &:=& d\psi +i\,\left(\,eA-\Gamma ^{\alpha\beta}\sigma
_{\alpha\beta}\right)\,\psi\,,\label{covder1}\\
\overline{D\psi} &:=&
d\overline{\psi}-i\,\overline{\psi}\,\left(\,eA-\Gamma
^{\alpha\beta}\sigma _{\alpha\beta}\right)\,.\label{covder2}
\end{eqnarray}
In analogy to the electromagnetic field strength (\ref{Fdef}), torsion (\ref{torsiondef}) and curvature (\ref{curvdef}) are to be regarded as the field strengths of
translations and of the Lorentz group respectively. The variations of all these objects are summarized as
\begin{eqnarray}
\delta \vartheta ^\alpha &=&-\,\vartheta ^{\,\beta}\beta _\beta {}^\alpha\,,\label{var-1}\\
\delta D\psi &=&\left(\,i\lambda +i\beta ^{\alpha\beta}\sigma
_{\alpha\beta}\,\right) D\psi\,,\label{var-2}\\
\delta\overline{D\psi}&=&-\,\overline{D\psi}\,\left(\,i\lambda
+i\beta ^{\alpha\beta}\sigma _{\alpha\beta}\,
\right)\,,\label{var-3}\\
\delta F &=& 0\,,\label{var-4}\\
\delta T^\alpha &=&- T^\beta \beta _\beta{}^\alpha \,,\label{var-5}\\
\delta R_\alpha{}^\beta &=& \beta _\alpha{}^\gamma R_\gamma{}^\beta
-\beta _\gamma{}^\beta R_\alpha{}^\gamma\,,\label{var-6}
\end{eqnarray}
calculated from (\ref{varcoordGoldstone})--(\ref{varlorconn}) applied to definitions (\ref{tetraddef})--(\ref{covder2}).

\section{Geometrical meaning of some dynamical variables}

Tetrads (\ref{tetrad}) (the pullback of (\ref{tetrad}) to the base space, in fact) can be chosen as a 1-form basis of the cotangent space. The corresponding {\it affine space} dual basis is taken to be the local reference frame $(\,o_x\,,e_\alpha\,)\,$, attached to each point $x$ of the base space \cite{Cartan} \cite{Gronwald:1997bx}, consisting of an origin $o_x\,$, together with a vector basis $e_\alpha$ defined by the condition $e_\alpha\rfloor\vartheta ^\beta =\delta _\alpha ^\beta\,$. Locally, the coordinate-like fields $\xi ^\alpha$ allow to define the position relative to a frame $(\,o_x\,,e_\alpha\,)\,$ as
\begin{equation}
p_x := o_x +\xi ^\alpha e_\alpha\,,\label{position}
\end{equation}
so that the $\xi ^\alpha$'s, although gauge-theoretical in origin, reveal to be interpretable as the components of a relative spacetime position vector. By taking
\begin{equation}
\delta o_x =\epsilon ^\alpha e_\alpha\,,\qquad \delta e_\alpha =\beta _\alpha{}^\beta \,e_\beta\,,\label{varorigin}
\end{equation}
together with (\ref{varcoordGoldstone}), the position (\ref{position}) results to be gauge invariant, that is
\begin{equation}
\delta p_x =0\,.\label{varposition}
\end{equation}
On the other hand, by introducing the translational and Lorentz connections, related respectively to the origin and to the vector basis as
\begin{equation}
\nabla o_x = {\buildrel (T)\over{\Gamma ^\alpha}}\otimes
e_\alpha\,,\qquad \nabla e_\alpha =\Gamma _\alpha{}^\beta\otimes
e_\beta\,,\label{connections}
\end{equation}
we find
\begin{equation}
\nabla p_x =\vartheta ^\alpha\otimes
e_\alpha\,,\label{displacement}
\end{equation}
with $\vartheta ^\alpha$ given by (\ref{tetrad}), showing the tetrad to originate from position transport. The line element $ds^2 :=o_{\alpha\beta}\,\vartheta ^\alpha\otimes\vartheta ^\beta\,$, built with the Minkowski metric and the tetrads (and equal to the standard Riemannian line element $ds^2 = g_{ij} dx^i dx^j\,$), can be regarded as a sort of (\ref{displacement}) squared. By acting again on (\ref{displacement}) one generates torsion
\begin{equation}
\nabla\nabla p_x = T^\alpha\otimes
e_\alpha\,,\label{nablators}
\end{equation}
while a double action on the basis vectors produces curvature
\begin{equation}
\nabla\nabla e_\alpha =R_\alpha{}^\beta\otimes
e_\beta\,.\label{nablacurv}
\end{equation}
Formulas (\ref{position})--(\ref{nablacurv}) provide a simple geometrical meaning to the dynamical objects $\xi ^\alpha\,$, ${\buildrel (T)\over{\Gamma ^\alpha}}\,$, and $\Gamma ^{\alpha\beta}\,$ in (\ref{constfields}), as much as to (\ref{tetraddef})--(\ref{curvdef}).

\section{On motion}

Let us notice that the structure (\ref{tetrad}) of tetrads, with the help of definition (\ref{position}) and (\ref{displacement}), allows to outline a mathematical description of motion in terms of the coordinate-like fields $\xi ^\alpha\,$. As a useful notational tool, we introduce besides ordinary Lie derivatives (\ref{Liederdef}) the covariant Lie derivatives \cite{Hehl:1995ue}, which for the particular case of the {\it time vector} $n$ are defined as
\begin{equation}
{\cal \L\/}_n\alpha ^A:=\,n\rfloor D\alpha ^A+
D\,(n\rfloor\alpha ^A\,)\,.\label{covLiederiv2}
\end{equation}
For instance
\begin{eqnarray}
{\cal \L\/}_n\vartheta ^\alpha := n\rfloor D\vartheta ^\alpha + D\left( n\rfloor\vartheta ^\alpha\right) =T_{\bot}^\alpha +D\vartheta _{\bot}^\alpha
\,.\label{thetaLiederiv}
\end{eqnarray}
Vertical gauge variations along fibers don't affect the position {\it points} (sections) $p_x$ due to their invariance (\ref{varposition}). However, when a horizontal displacement occurs between neighboring fibers from position $p_x$ to $p_{x+dx}\,$ along a worldline (that is, along a path parametrized by $\tau$ having $n$ as its tangent vector), then according to (\ref{displacement})
\begin{equation}
\nabla _n\,p_x =\nabla _n\,\left(\,o_x +\xi ^\alpha e_\alpha\,\right) = \vartheta _{\bot}^\alpha\otimes e_\alpha\,,\label{positmot}
\end{equation}
with the quantity
\begin{equation}
\vartheta _{\bot}^\alpha := {\cal \L\/}_n\,\xi ^\alpha + {\buildrel (T)\over{\Gamma _{\bot}^\alpha}}\,,\label{motion}
\end{equation}
acting as the covariant four-velocity. As read out from (\ref{motion}), horizontal displacements cause the measurable relative position vector $\xi ^\alpha$ in (\ref{position}) to evolve with respect to {\it parametric time}, while a contribution due to the change of origin ensures covariance. From (\ref{motion}) we get the covariant acceleration
\begin{eqnarray}
{\cal \L\/}_n\,\vartheta _{\bot}^\alpha &=& {\it{l}}_n\,\vartheta _{\bot}^\alpha +\Gamma  _{\bot\beta}{}^\alpha\,\vartheta _{\bot}^\beta \nonumber\\
&=& {\cal \L\/}_n\,{\cal \L\/}_n\,\xi ^\alpha + {\cal \L\/}_n\,{\buildrel (T)\over{\Gamma _{\bot}^\alpha}}\,,\label{acceleration}
\end{eqnarray}
including a sort of force contribution associated to the origin. (In principle, it should be possible to reexpress {\it parametric time} evolution as evolution with respect to clock time, say to $\xi ^0\,$.)

Einstein's general relativistic geodesic equations for classical test particles establish the vanishing of (\ref{acceleration}). To get such a simple equation from (\ref{sigmamattconserv}), we have to consider phenomenological matter, for instance that described by a dust model with matter currents $J=0\,$, $\tau _{\alpha\beta}=0\,$ and $\Sigma ^{\rm matt}_\alpha =\rho\,\vartheta _{\bot\alpha}\,\vartheta _{\bot}^\beta\,\eta _\beta\,$, being $\rho$ a flow density 0-form. Then, taking into account, as derived from (\ref{antisym2form}) and (\ref{antisym3form}) respectively, that $\eta _{\bot\alpha} =-\,{}^\#\underline{\vartheta}_\alpha\,$ and $\underline{\eta}_\alpha =-\,{}^\#\vartheta _{\bot\alpha}\,$, and on the other hand $\eta _{\bot\alpha\beta} =\,{}^\#(\,\underline{\vartheta}_\alpha\wedge\underline{\vartheta}_\beta\,)\,$ and $\underline{\eta}_\alpha =-\,{}^\#(\,\vartheta _{\bot\alpha}\underline{\vartheta}_\beta -\vartheta _{\bot\beta}\underline{\vartheta}_\alpha\,)\,$, eq. (\ref{sigmamattconserv}) yields
\begin{equation}
{\cal \L\/}_n\,\left(\,\vartheta _{\bot\alpha}\,{}^\#\rho\,\right) = -\rho\,\,\,{}^\#\underline{\vartheta}_\alpha\wedge \vartheta _{\bot\beta}\, T_{\bot}^\beta\,,\label{Einstmot}
\end{equation}
which for vanishing torsion and ${\it{l}}_n\,{}^\#\rho =0\,$ reproduces the desired result ${\cal \L\/}_n\,\vartheta _{\bot}^\alpha = 0\,$. Fundamental matter gives rise to more complicated equations involving ${\cal \L\/}_n\,\vartheta _{\bot}^\alpha\,$ by using either (\ref{sigmamattconserv}) with (\ref{transversmattenmom1}), or (\ref{sigmaemconserv}) with (\ref{transversemenmom1}), or (\ref{covfieldeq2}) with the transversal part of the total energy-momentum
\begin{eqnarray}
\underline{\Pi}_{\alpha} &=& n\rfloor \Bigl\{\,\vartheta _ \alpha\wedge\left(\,\epsilon + D\vartheta ^\beta _{\bot}\wedge H_\beta\right)\nonumber\\ &&\hskip0.5cm -2\,\vartheta ^\beta _{\bot} D\left(\,\tau _{\alpha\beta} + {a_0\over{2\kappa}}\,D\eta _{\alpha\beta}\,\right)\,\Bigr\}\,,\label{transverstotenmom}
\end{eqnarray}
(identical with $\,{}^{\#}\pi ^{\xi}_\alpha$ in (\ref{moms1})\,), resulting from putting together (\ref{transversmattenmom1}), (\ref{transversemenmom1}) and (\ref{transversgrenmom}), the latter one evaluated for (\ref{gravlagr}), with (\ref{momentdecomp}) and (\ref{energydec}).
Let us conclude claiming that there are the dynamical relative positions $\xi ^\alpha$ involved in the field equations, rather than the underlying quite metaphysical base space points (or their coordinates), that describe observable spacetime.

\section{Consequences of the base space foliation}

The foliation of the base space considered by us rests on the introduction of a time-like vector field $n=n^i\partial _i\,$, tangent to a congruence of worldlines, whose direction is fixed with respect to the 1-form $N d\tau$ by requiring both to satisfy the condition $n\rfloor ( N d\tau\,)=1\,$. In terms of the lapse $N$ and the shift $N^a$ functions, it is possible to rewrite the {\it parametric time vector field} as $\,n={1\over N}\,(\partial _\tau -N^a\partial _a\,)\,$, with $\partial _a$ as space derivatives. In the present appendix we extend the decomposition (\ref{foliat1}) of any $p$-form into longitudinal and transversal parts (\ref{long-part}) and (\ref{trans-part}) respectively, to both, exterior derivatives of forms and Hodge dual forms (\ref{dualform}), and then we present a foliated version of the variations presented in Section III A.

In analogy to (\ref{foliat1}), exterior derivatives decompose as
\begin{equation}
d\,\alpha = N d\tau\wedge\left( d\,\alpha\right)_{\bot}
+\underline{d\,\alpha }\,,\label{derivfoliat}
\end{equation}
with
\begin{equation}
\left( d\,\alpha\right)_{\bot} = {\it{l}}_n\alpha  - d\,\alpha_{\bot} ={\it{l}}_n\underline{\alpha}
-{1\over N}\,\underline{d}\,\bigl(\,N\alpha
_{\bot}\bigr)\,.\label{longitderiv}
\end{equation}
On the other hand, the Hodge dual of an arbitrary $p$-form $\alpha$, as defined by (\ref{dualform}), decomposes as
\begin{equation}
{}^*\alpha =\,(-1)^p\,N d\tau\wedge {}^{\#}\underline{\alpha} -
{}^{\#}\alpha _{\bot}\,,\label{foliat2}
\end{equation}
being $^\#$ the Hodge dual operator in the 3-dimensional spatial sheets. Taking (\ref{foliat2}) into account, we derive the following results, which are useful to reproduce the calculations of the main text. In the four--dimensional spacetime with Lorentzian signature, the double application of the Hodge dual operator reproduces $\alpha$ itself up to the sign as $^{**}\alpha
=-(-1)^p\,\alpha$. From this relation we deduce
\begin{eqnarray}
^{\#\#}\alpha _{\bot}&=&\alpha _{\bot}\,,\label{doublestar1}\\
\,^{\#\#}\underline{\alpha}&=&\underline{\alpha}\,.\label{doublestar2}
\end{eqnarray}
On the other hand, from $\vartheta ^\alpha\wedge e_\alpha\rfloor\alpha =\,p\,\alpha$
we find
\begin{eqnarray}
\underline{\vartheta}^\alpha\wedge e_\alpha\rfloor\alpha
_{\bot}&=&(\,p-1\,)\,\alpha _{\bot}\,,\label{intp1}\\
\underline{\vartheta}^\alpha\wedge
e_\alpha\rfloor\underline{\alpha}
&=&\,p\,\,\underline{\alpha}\,,\label{intp2}
\end{eqnarray}
and from the further relation $^*(\alpha\wedge\vartheta _\alpha\,)= e_\alpha\rfloor\,^*\alpha \,$ involving Hodge duality we get
\begin{eqnarray}
^\#(\alpha _{\bot}\wedge\underline{\vartheta}_\alpha ) &=& e_\alpha\rfloor\,^\#\alpha _{\bot}\,,\label{3dHodge1}\\
{}^\#(\underline{\alpha}\wedge\underline{\vartheta}_\alpha ) &=& e_\alpha\rfloor\,^\#\underline{\alpha}
\,.\label{3dHodge2}
\end{eqnarray}
Furthermore, being $\alpha$ and $\beta$ differential forms of the same degree $p$, equation $\alpha\wedge{}\,^*\beta =\beta \wedge{}\,^*\alpha$ holds, yielding
\begin{eqnarray}
\alpha _{\bot}\wedge\,{}^\#\beta _{\bot} &=&\beta
_{\bot}\wedge\,{}^\#\alpha _{\bot}\,,\label{crossinterch1}\\
\underline{\alpha}\wedge\,{}^\#\underline{\beta}
&=&\underline{\beta}\wedge\,{}^\#\underline{\alpha}\,.\label{crossinterch2}
\end{eqnarray}
We end this collection of equations by reformulating variations (\ref{varlag1}) and (\ref{varlag4}) on a foliated base space respectively as
\begin{eqnarray}
\delta L &=& N d\tau\wedge\delta L_{\bot} = N d\tau\wedge\Bigl[
\,\delta Q_{\bot}\wedge {{\partial L_{\bot}}\over{\partial
Q_{\bot}}}+\delta \underline{Q}\wedge {{\partial
L_{\bot}}\over{\partial\underline{Q}}}\nonumber\\
&&\hskip1.1cm +\delta \left( dQ\right)_{\bot}\wedge {{\partial L_{\bot}}\over{\partial
{\it{l}}_n\underline{Q}}}+\delta\underline{dQ}\wedge {{\partial
L_{\bot}}\over{\partial\underline{dQ}}}\,\Bigr]\,,\label{varlag3}
\end{eqnarray}
and
\begin{eqnarray}
\delta L &=& N d\tau\wedge\left(\,\delta Q_{\bot}\wedge {{\delta L_{\bot}}\over{\delta Q_{\bot}}} +\delta\underline{Q}\wedge {{\delta L_{\bot}}\over{\delta\underline{Q}}}\,\right)\nonumber\\
&&+ d \Bigl[\,N d\tau\wedge\left( \delta Q_{\bot}\wedge
{{\partial L_{\bot}}\over{\partial
{\it{l}}_n\underline{Q}}}-\delta \underline{Q}\wedge {{\partial
L_{\bot}}\over{\partial\underline{dQ}}}\,\right)\nonumber\\
&&\hskip1.0cm + \delta\underline{Q}\wedge {{\partial L_{\bot}}\over{\partial
{\it{l}}_n\underline{Q}}}\,\Bigr]
\,,\label{varlag5}
\end{eqnarray}
in terms of the foliated Lagrangian density form $L =N d\tau\wedge L_{\bot}\,$, depending on the longitudinal and transversal parts of the dynamical variables $Q = N d\tau\wedge Q_{\bot} + \underline{Q}\,$. We find the consistence requirements
\begin{eqnarray}
{{\partial L}\over{\partial Q}}&=&(-1)^p\,N d\tau\wedge{{\partial
L_{\bot}}\over{\partial\underline{Q}}}+{{\partial
L_{\bot}}\over{\partial Q_{\bot}}}\,,\label{condit1}\\
{{\partial L}\over{\partial dQ}} &=&-(-1)^p\,N d\tau\wedge
{{\partial L_{\bot}}\over{\partial\underline{dQ}}} +{{\partial
L_{\bot}}\over{\partial {\it{l}}_n\underline{Q}}}
\,.\label{condit2}
\end{eqnarray}
The field equations are decomposed as
\begin{equation}
{{\delta L}\over{\delta Q}} = (-1)^p\,N d\tau\wedge\,{{\delta L_{\bot}}\over{\delta\underline{Q}}} +{{\delta L_{\bot}}\over{\delta Q_{\bot}}}\,,\label{foliatfieldeq0}
\end{equation}
with
\begin{eqnarray}
{{\delta L_{\bot}}\over{\delta Q_{\bot}}}&:=& {{\partial
L_{\bot}}\over{\partial Q_{\bot}}}-(-1)^p\,\,\underline{d}\,\left(
{{\partial L_{\bot}}\over{\partial
{\it{l}}_n\underline{Q}}}\right)\,,\label{foliatfieldeq1}\\
{{\delta L_{\bot}}\over{\delta\underline{Q}}}&:=& {{\partial
L_{\bot}}\over{\partial\underline{Q}}} -{\it{l}}_n \left(
{{\partial L_{\bot}}\over{\partial
{\it{l}}_n\underline{Q}}}\right)-(-1)^p\,{1\over
N}\,\underline{d}\,\left( N {{\partial
L_{\bot}}\over{\partial\underline{dQ}}}\right)
\,,\nonumber\\
\label{foliatfieldeq2}
\end{eqnarray}
standing $p$ for the degree of the $p$-form $Q$.

\section{Effect of the foliation on tetrads}

Tetrads are of particular relevance for the present work. Their foliation decomposition (\ref{foliat1}) reads
\begin{equation}
\vartheta ^\alpha = N d\tau\,\vartheta ^\alpha _{\bot} +
\underline{\vartheta}^\alpha\,,\label{tetradfoliat}
\end{equation}
being possible to express the {\it parametric time vector} $n=n^i\partial _i$ alternatively in terms of the longitudinal part $\vartheta ^\alpha _{\bot}:=n\rfloor\vartheta ^\alpha $ as $n =\vartheta ^\alpha _{\bot} e_\alpha$. The Hodge dual (\ref{antisym3form}) of the tetrad decomposes according to (\ref{foliat2}) as
\begin{equation}
\eta ^\alpha := {}^*\vartheta ^\alpha = -N
d\tau\,{}^{\#}\underline{\vartheta}^\alpha - \,{}^{\#}\vartheta
^\alpha _{\bot}\,.\label{Hodgetetradfoliat}
\end{equation}
On the other hand we know \cite{Hehl:1995ue} that $\vartheta ^\alpha\wedge\eta _\beta = \delta ^\alpha _\beta\,\eta\,$, so that with the help of (\ref{eta4form}) and (\ref{foliat2}) we find
\begin{equation}
\vartheta ^\alpha\wedge\eta _\alpha = 4\,N d\tau\wedge\,{}^{\#}1
\,,\label{volelement2}
\end{equation}
while from (\ref{tetradfoliat}) and (\ref{Hodgetetradfoliat}) we get
\begin{equation}
\vartheta ^\alpha\wedge\eta _\alpha = N d\tau\wedge\left(
-\vartheta ^\alpha _{\bot}\,{}^{\#}\vartheta _{\bot\alpha}
+\underline{\vartheta}^\alpha\wedge
{}^{\#}{\underline{\vartheta}\,}_{\alpha}\,\right)\,.\label{volelement1}
\end{equation}
Comparison of (\ref{volelement2}) and (\ref{volelement1}) yields
\begin{equation}
-\vartheta ^\alpha _{\bot}\,{}^{\#}\vartheta _{\bot\alpha}
+\underline{\vartheta}^\alpha\wedge
{}^{\#}{\underline{\vartheta}\,}_{\alpha} =\,{}^{\#}4
\,.\label{volelement3}
\end{equation}
Now, from (\ref{antisym3form}) with (\ref{tetradfoliat}) we find the explicit form of the longitudinal part of (\ref{Hodgetetradfoliat}) to be
\begin{equation}
{}^{\#}{\underline{\vartheta}\,}_{\alpha}
=\,{1\over{2}}\,\vartheta
_{\bot}^\mu\,\eta
_{\mu\alpha\beta\gamma}\,\underline{\vartheta}^\beta
\wedge\underline{\vartheta}^\gamma\,,\label{transtetrad}
\end{equation}
so that on the one hand we get trivially
\begin{equation}
\vartheta ^\alpha _{\bot}\,
{}^{\#}{\underline{\vartheta}\,}_{\alpha}=0 \,,\label{orthog}
\end{equation}
and on the other hand
\begin{equation}
\underline{\vartheta}^\alpha\wedge
{}^{\#}{\underline{\vartheta}\,}_{\alpha}
=\,3\,{1\over{3!}}\,\vartheta _{\bot}^\mu\,\eta
_{\mu\alpha\beta\gamma}\,
\underline{\vartheta}^\alpha\wedge\underline{\vartheta}^\beta
\wedge\underline{\vartheta}^\gamma
=\,{}^{\#}3 \,.\label{3volum}
\end{equation}
Putting together (\ref{volelement3}), (\ref{orthog}) and (\ref{3volum}), and taking the Hodge dual when it simplifies the expressions, we finally get
\begin{equation}
\vartheta ^\alpha
_{\bot}\,{\underline{\vartheta}\,}_{\alpha}=0\,,\quad \vartheta
_{\bot}^\alpha\,\vartheta _{\bot\alpha} =-1\,,\quad
\underline{\vartheta}^\alpha\wedge\,{}^{\#}\underline{\vartheta}_\alpha
=\,{}^{\#}3\,.\label{quadrats}
\end{equation}
Since $n =\vartheta ^\alpha _{\bot}\,e_\alpha\,$, being $\vartheta _{\bot}^\alpha\,\vartheta _{\bot\alpha} =-1\,$, we formally conclude that $N d\tau =-\vartheta _\alpha \vartheta ^\alpha _{\bot}\,$, so that $e_\alpha\rfloor (N d\tau ) =-\vartheta _{\bot\alpha}\,$.

\section{Universal coupling of translational variables}

Extensive use is made in the main text of the eta basis, consisting in the Hodge dual of exterior products of tetrads. With the help of the Levi-Civita object
\begin{equation}
\eta ^{\alpha\beta\gamma\delta}:=\,^*(\vartheta
^\alpha\wedge\vartheta ^\beta\wedge\vartheta
^\gamma\wedge\vartheta ^ \delta\,)\,,\label{levicivita}
\end{equation}
we define
\begin{eqnarray}
\eta ^{\alpha\beta\gamma}&:=&\,^*(\vartheta ^\alpha\wedge\vartheta
^\beta\wedge\vartheta ^\gamma\,)=\,\eta
^{\alpha\beta\gamma}{}_\delta\,\vartheta ^
\delta\,,\label{antisym1form}\\
\eta ^{\alpha\beta}&:=&\,^*(\vartheta ^\alpha\wedge\vartheta
^\beta\,)={1\over{2!}}\,\eta ^{\alpha\beta}{}_{\gamma\delta}
\,\vartheta ^\gamma\wedge\vartheta ^
\delta\,,\label{antisym2form}\\
\eta ^\alpha &:=&\,^*\vartheta ^\alpha ={1\over{3!}}\,\eta
^\alpha{}_{\beta\gamma\delta} \,\vartheta ^\beta\wedge\vartheta
^\gamma\wedge\vartheta ^ \delta\,,\label{antisym3form}\\
\eta &:=&\,^*1 ={1\over{4!}}\,\eta
_{\alpha\beta\gamma\delta}\,\vartheta ^\alpha\wedge\vartheta
^\beta\wedge\vartheta ^\gamma\wedge\vartheta ^
\delta\,,\label{eta4form}
\end{eqnarray}
where (\ref{eta4form}) is the four--dimensional volume element. Being tetrads $\vartheta ^\alpha$ chosen as a basis of the cotangent space, an arbitrary $p$-form $\alpha$ reads
\begin{equation}
\alpha ={1\over{p\,!}}\,\vartheta ^{\alpha _1}\wedge
...\wedge\vartheta ^{\alpha _p}\,(e_{\alpha _p}\rfloor ...
e_{\alpha _1}\rfloor\alpha\,)\,,\label{pform}
\end{equation}
while its Hodge dual becomes expressed in terms of the eta basis (\ref{levicivita})--(\ref{eta4form}) as
\begin{equation}
\,{}^*\alpha ={1\over{p\,!}}\,\eta ^{\alpha _1 ... \alpha
_p}\,(e_{\alpha _p}\rfloor ... e_{\alpha
_1}\rfloor\alpha\,)\,.\label{dualform}
\end{equation}
Notice that comparison of the variations of (\ref{pform}) and (\ref{dualform}) to each other yields the relation
\begin{equation}
\delta \,{}^*\alpha =\,{}^*\delta\alpha -{}^*\left(\delta
\vartheta ^\alpha\wedge e_\alpha\rfloor\alpha\,\right)
+\delta\vartheta ^\alpha\wedge\left( e_\alpha\rfloor
{}^*\alpha\,\right)\,,\label{dualvar}
\end{equation}
which has a decisive relevance in showing that the variation of forms affected by the Hodge star operator involve variation of the tetrads (\ref{tetrad}) and thus of $\xi ^\alpha$ and ${\buildrel (T)\over{\Gamma ^\alpha}}\,$. It is through the coupling of the translational variables to any other quantity that the universal influence of gravity on other fields takes place. This fact reveals itself in dynamics in the existence of contributions to the total energy-momentum (\ref{momentdecomp}) arising from any matter or force parts of the Lagrangian. The foliated version of (\ref{antisym1form})--(\ref{eta4form}), as much as of (\ref{dualvar}), are relevant for the complete Hamiltonian approach to be published elsewhere.

\end{document}